%% file: main.tex
\documentclass[acmsmall,screen,nonacm]{acmart}

\input{utils/packages}
\input{utils/commands}

\input{utils/copyright}

\begin{document}

\title{Transpiler Autotuning with Predictive Models for Quantum~Circuit~Optimization}
\input{utils/authors}
\renewcommand{\shortauthors}{Malkowski et al.}

\begin{abstract}
\input{sections/00_abstract}
\end{abstract}

\maketitle

\input{sections/01_introduction}

\input{sections/02_background}
x\input{sections/03_overview}
\input{sections/04_concept}
\input{sections/05_evaluation}

\input{sections/06_related_work}
\input{sections/07_conclusion}

\begin{acks}
This work was funded by the German Federal Ministry of Research, Technology and Space (BMFTR) in the project QuSol (Grant No. 13N17170) and the German Research Foundation (DFG) in the project MoQel (Grant No. SCHA 1635/20-1).

We thank IBM and Marcel Pfaffhauser for hosting the IBM Hackathon in Munich, where the idea for this work originated. We further thank Robert Wille for encouraging us to extend the Hackathon idea into a full paper, and Sebastian Krieter for his help with YASA.
\end{acks}

\bibliographystyle{ACM-Reference-Format}
\bibliography{bib}

\end{document}

%% file: utils/packages.tex
\usepackage{algorithmic}
\usepackage{graphicx}
\usepackage{textcomp}
\usepackage{hyperref}
\usepackage{xcolor}
\usepackage{listings}
\usepackage{float}
\usepackage{booktabs}
\usepackage{braket}
\usepackage[caption=false,font=footnotesize]{subfig}

\AtBeginDocument{%
  }

\usepackage{glossaries-extra}
\setabbreviationstyle[acronym]{long-short}
\glssetcategoryattribute{acronym}{nohyperfirst}{true}

%% file: utils/commands.tex
\theoremstyle{definition}
\newtheorem{example}{Example}[section]
\newtheorem{definition}{Definition}[section]

\newacronym{nisq}{NISQ}{Noisy Intermediate-Scale Quantum}
\newacronym{qpu}{QPU}{Quantum Processing Unit}
\newacronym{qaoa}{QAOA}{Quantum Approximate Optimization Algorithm}
\newacronym{vqe}{VQE}{Variational Quantum Eigensolver}
\newacronym{qnn}{QNN}{Quantum Neural Network}
\newacronym{shap}{SHAP}{SHapley Additive exPlanations}
\newacronym{dag}{DAG}{Directed Acyclic Graph}
\newacronym{isa}{ISA}{Instruction Set Architecture}
\newacronym{ndcg}{NDCG}{Normalized Discounted Cumulative Gain}

\newcommand{\mqtBench}{MQT Bench}
\newcommand{\mqtPred}{MQT Predictor}

\newcommand{\learningToRank}{learning-to-rank}
\newcommand{\explVars}{explanatory variables}
\newcommand{\execOrder}{execution order}
\newcommand{\strategyRule}{strategy rule}
\newcommand{\dependencyChainRule}{dependency chain rule}
\newcommand{\mandatoryTranslationRule}{mandatory translation rule}

%% file: utils/copyright.tex
\setcopyright{acmlicensed}
\copyrightyear{2018}
\acmYear{2018}
\acmDOI{XXXXXXX.XXXXXXX}
\acmConference[Conference acronym 'XX]{Make sure to enter the correct
  conference title from your rights confirmation email}{June 03--05,
  2018}{Woodstock, NY}
\acmISBN{978-1-4503-XXXX-X/2018/06}

%% file: utils/authors.tex
\author{Piotr Malkowski}
\email{piotr.malkowski@kit.edu}
\orcid{0009-0004-7026-0016}
\correspondingauthor
\affiliation{%
  \institution{Karlsruhe Institute of Technology}
  \country{Germany}
}

\author{Domenik Eichhorn}
\email{domenik.eichhorn@kit.edu}
\orcid{0000-0001-9428-024X}
\correspondingauthor
\affiliation{%
  \institution{Karlsruhe Institute of Technology}
  \country{Germany}
}

\author{Joshua Ammermann}
\email{joshua.ammermann@kit.edu}
\orcid{0000-0001-5533-7274}
\affiliation{%
  \institution{Karlsruhe Institute of Technology}
  \country{Germany}
}

\author{Rinor Kelmendi}
\email{rinor.kelmendi@kit.edu}
\orcid{0009-0003-0951-6871}
\affiliation{%
  \institution{Karlsruhe Institute of Technology}
  \country{Germany}
}

\author{Nick Poser}
\email{nick.poser@student.kit.edu}
\orcid{0009-0001-1915-4419}
\affiliation{%
  \institution{Karlsruhe Institute of Technology}
  \country{Germany}
}

\author{Patrick Hopf}
\email{patrick.hopf@tum.edu}
\orcid{0009-0008-1358-2501}
\affiliation{%
  \institution{Technical University of Munich}
  \country{Germany}
}

\author{Ina Schaefer}
\email{ina.schaefer@kit.edu}
\orcid{0000-0002-7153-761X}
\affiliation{%
  \institution{Karlsruhe Institute of Technology}
  \country{Germany}
}

%% file: sections/00_abstract.tex
Quantum software engineering is an emerging research field focusing on efficiently embedding the quantum programming paradigm into existing software ecosystems.
A key aspect of this field is the realization of quantum algorithms using gate-based programming and the subsequent low-level optimization of the resulting quantum circuits, a process that is commonly performed by so-called transpilation pipelines.
One significant challenge in these pipelines is determining which optimizations to apply to a given circuit. This decision is usually based on fixed default configurations that are uniformly applied to all circuits, frequently resulting in missed opportunities for more aggressive circuit optimization.
In this work, we tackle this challenge by applying autotuning with supervised machine learning to develop an automated method for selection of transpiler passes.
To train our machine-learning models, we employ feature-model based sampling to generate a representative dataset that examines how different combinations of Qiskit transpiler passes perform across thousands of circuits drawn from the state-of-the-art benchmarking suite MQT Bench.
Using these data, we build a predictive model extension for the Qiskit transpilation pipeline that uses a machine learning model to automatically select combinations of transpiler passes aiming to achieve a maximum reduction in two-qubit gates.
Our empirical evaluation shows that the combinations selected by our model are never outperformed by Qiskit's optimization levels, achieve on average an additional 19.1$\%$ - 32.4$\%$ reduction in two-qubit gates, and for some circuits finds reductions of up to $95.8\%$ in cases where Qiskit achieves no reduction at all.

%% file: sections/01_introduction.tex
\section{Introduction}

Quantum computing represents a novel paradigm that can strongly impact the field of modern computer science by providing up to superpolynomial speedups for well-structured problems~\cite{aaronson2022much}. 
In the current \gls{nisq} era, early quantum devices with relatively few qubits are available, but they are not yet mature enough to be scaled to deliver advantages in practical, real-world applications~\cite{shaydulin2024evidence}.
A key challenge in quantum computing is the creation of new software technologies that can embed this emerging quantum paradigm into existing classical infrastructures, giving us the potential to exploit these up to superpolynomial speedups in practice once major advances in quantum hardware are achieved~\cite{desdentado2025quantum}.
To prevent a quantum software crisis analogous to the classical software crisis of the 1960s, the systematic development of quantum software has emerged as a rapidly evolving research area~\cite{moguel2020roadmap}.

A central concern in quantum software engineering is the design of programming languages and libraries that can accommodate this novel technological landscape.
One prominent example of a quantum software framework is Qiskit, a language and ecosystem developed by IBM~\cite{Qiskit}, which is currently considered as the most influential quantum programming framework~\cite{upadhyay2025analyzing}. 
Qiskit lowers the barrier to entry for quantum computing by enabling users to construct and deploy \textit{quantum circuits}, thereby expressing the manipulation of quantum mechanical systems via a gate-based programming model~\cite{kwon2021gate}. 
Quantum circuits are analogous to classical logic circuits, consisting of a sequence of \textit{quantum gates} that are applied to qubits~\cite{hidary2021quantum}. 
They constitute the fundamental representation used to instruct \glspl{qpu} to perform computational tasks.

Similarly to compiling for a CPU, compiling for a specific \gls{qpu} requires consideration of the device's characteristics regarding specific figures of merit, such as circuit depth and the number of available gates~\cite{hopfImprovingFiguresMerit2025}.
Currently, the majority of quantum circuit compilation is performed by transpilation pipelines provided by frameworks like Qiskit~\cite{Qiskit} and Tket~\cite{tket}. These frameworks provide a large set of so-called \textit{transpiler passes}, which modify circuits so that they match the characteristics required by a particular quantum device, or adjust them to meet pre-defined optimization goals, such as the reduction of two-qubit gates, which is the primary source of errors on \gls{nisq} devices~\cite{Preskill2018quantumcomputingin}.

A recurring challenge of quantum circuit transpilation is that it is typically not know in advance which configuration of transpilation passes will yield the best outcome. As a result, frameworks such as Qiskit~\cite{Qiskit} and Tket~\cite{tket} generally rely on predefined optimization levels, each corresponding to a progressively more resource-intensive, fixed set of optimization passes.  
Although this strategy is pragmatic, it often fails to exploit opportunities for more aggressive optimization, leading to a less efficient use of costly quantum computing resources.  
Addressing this challenge requires understanding which combinations of transpilation passes are likely to be most effective for a specific quantum circuit instance. However, this is difficult to decide because the large number of available passes creates a combinatorial explosion in the search space.  
To overcome this, we must devise a systematic approach that supports efficient exploration of this search space, allowing a circuit-aware selection of optimization passes tailored to different circuit classes, thus replacing today’s fixed optimization levels. 

In this work, we tackle this challenge by presenting a methodology that leverages autotuning~\cite{Basu2013TowardsMA} to build a machine learning predictor capable of reliably recommending  combinations of transpiler passes for a given circuit, with the goal of maximizing the reduction in two-qubit gates.
This paper describes the multiple steps that were necessary to implement this methodology within a real-world quantum computing framework (Qiskit). 
In addition, we provide detailed evaluations demonstrating that our approach can reliably surpass the existing optimization levels of the Qiskit transpilation pipeline regarding hardware-independent optimizations, and that we are competitive with other state-of-the-art machine learning methods such as the \mqtPred{}~\cite{quetschlich2024mqt_predictor}. 
To realize our methodology, we make the following three contributions:
(1) We construct a novel data set that allows us to analyze the behavior and efficiency of transpilation passes across various classes of quantum algorithms, including the \gls{qaoa}~\cite{farhi2014quantumapproximateoptimizationalgorithm}, the \gls{vqe}~\cite{kandala2017hardware}, and \glspl{qnn}~\cite{kwak2021quantumneuralnetworksconcepts}. To build this data set, we draw 1,943 quantum circuit instances from the well-established \mqtBench{}~\cite{quetschlich2023mqtbench} library and then examine how the optimization passes available in the Qiskit perform on them. To manage the combinatorial blow-up inherent in this setting, we design a feature model~\cite{kang1990feature} that captures the configurability of the Qiskit transpilation pipeline and apply \textit{3}-wise interaction sampling~\cite{krieter_yasa_2020} to derive a representative subset of 62 combinations of transpilation passes. We then run these 62 configurations on all 1,943 circuits, creating a large scale data set that can be used to analyze transpilation passes.
(2) We use this newly constructed data set to train a state-of-the-art machine learning model using XGBoost~\cite{xgboost-paper} with a \learningToRank{} approach~\cite{learning-to-rank-paper-nair}. For a new, unseen circuit, this predictive model then serves as a fast surrogate by predicting the configuration that is expected to yield the largest reduction in two-qubit gates.
(3) We determine which combinations of transpilation passes are helpful, harmful, or neutral for each distinct quantum circuit problem class by applying a \gls{shap} analysis~\cite{shap} to our predictive model. From this, we identify transpilation passes that, on average, provide the greatest benefit across all problem classes, as well as those that are especially effective for particular classes. These findings can influence how end-users approach their own quantum compilation tasks and how quantum transpilation frameworks design heuristic strategies to maximize optimization potential.

%% file: sections/02_background.tex
\section{Background}

This section provides background information on quantum compilation with a strong focus on the transpilation pipeline of Qiskit, which we improve later in this work. Furthermore, we give a general overview of compiler autotuning approaches and supervised machine learning, which are concepts on which we later base our methodology.

\subsection{Quantum Compilation and the Qiskit Compiler Pipeline}
\label{background:qiskit-subsec}

In the current \gls{nisq} era, quantum algorithms are typically expressed using a gate-based programming model~\cite{kwon2021gate} and implemented in domain-specific quantum programming languages such as Q\#~\cite{svore2018q}, or through quantum computing frameworks that extend existing languages, such as  Qiskit~\cite{Qiskit} and Qrisp~\cite{seidel2024qrisp}, which support gate-based programming via Python and similar interfaces. Similar to classical computer architectures, these gate-based quantum circuits rely on compilers to translate high-level program descriptions into an instruction set that can be executed on a QPU. A key challenge in quantum computing is that current quantum compilers must perform extensive circuit optimizations to adapt circuits for hardware devices that offer only a small number of noisy qubits with restricted connectivity. Consequently, a central component of any quantum compiler pipeline is circuit transpilation, a subroutine that transforms and optimizes circuits so they are compatible with a given quantum device and use fewer error-prone gates, such as 2-qubit gates.

\begin{figure}[h]
    \centering
    \includegraphics[width=1.0\linewidth]{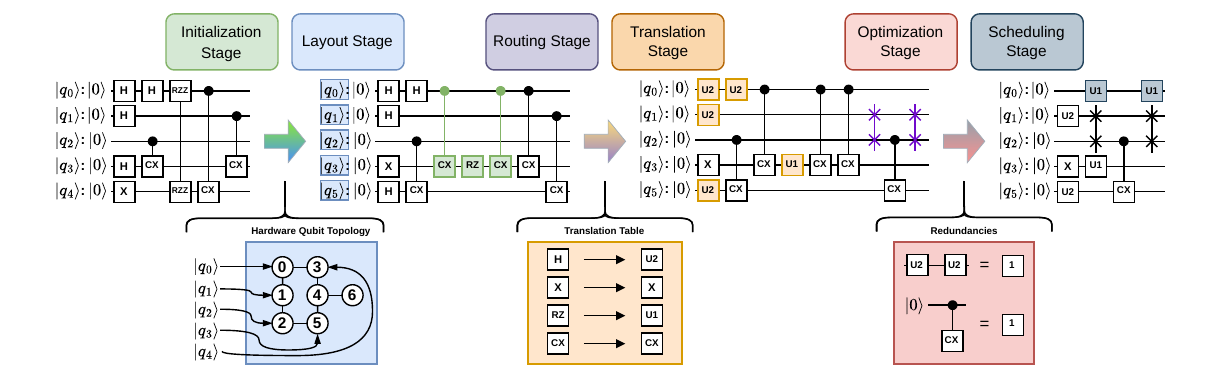}
    \caption{Overview of Qiskit's transpiler pipeline that illustrates how every stage may transform a given circuit.
    }
    \Description{Shown is figure that gives an overview of the Qiskit Transpilation Pipeline using an example circuit. In the Initialization Stage, the custom gate RZZ gate is decomposed into a CX, RZ and CX gate. The Layout stage maps the the qubits onto the topology of the circuit, so that the qubits with the most interaction are next to each other. Afterwards in the routing stage additional SWAP-Gates are used to preserve the connection between the qubits, which could not be mapped onto the topology. In the translation stage, the used gates are translated into ISA of the quantum hardware. The H-gate is translated into a U2 gate and the RZ Gate to the U1 gate with the proper parameters. The optimization stage reduces the redundancy and tries to optimize the circuit by deleting for example both U2 gates on $\ket{q_0}$ which nullify themselves. In the scheduling stage, additional gates, which do not affect the results, are implemented to reduce the risk of errors.}
    \label{fig:qiskit-transpiler-pipeline}
\end{figure}

In this paper, our aim is to improve the optimizations that a transpilation pipeline can perform by proposing a concept for the automatic selection of transpilation passes.
To realize this concept, we extend the existing quantum programming framework Qiskit~\cite{Qiskit}.
The Qiskit transpilation pipeline is divided into six stages, where each stage comprises several transpilation passes that can be combined to carry out transpilation.
Figure~\ref{fig:qiskit-transpiler-pipeline} shows a visual overview of this transpilation pipeline.

Within Qiskit's transpilation pipeline, the stages serve two primary purposes: (1) adapting a circuit to a specific device through layout, routing, scheduling, and translation stages, and (2) optimizing circuits, most notably by decreasing the number of two qubit gates in the init and optimization stages, which are the main contributors to error on \gls{nisq} devices. The init-stage, for example, carries out logical optimizations on abstract circuits and breaks down multi-qubit and custom gates into one- and two-qubit operations. This decomposition is necessary because most layout and routing algorithms are based on these gates. In this work, we will set a focus on transpilation passes from the init-stage, which can be grouped into two basic categories:

\begin{itemize}
    \item \textbf{Analysis passes}, which traverse the circuit’s \gls{dag} representation to infer properties, such as commutation relations and block collections, without altering the original circuit. The results of these analyses are stored in a shared property set.  
    \item \textbf{Transformation passes}, which operate on the \gls{dag} to improve the circuit structure. They typically depend on the property set generated by preceding analysis passes. A key aspect to emphasize is that the information stored in the property set is, however, not reused across multiple transformation passes. Instead, the required analysis passes are hard-coded and re-executed within each transformation pass \cite{javadi2024quantum}.
\end{itemize}

\subsection{Compiler Autotuning and Supervised Machine Learning}
\label{subsec:bacckground_compiler_autotuning_and_supervised_ml}

To avoid the manual trial-and-error exploration of different configurations in transpilation pipelines, autotuning~\cite{Basu2013TowardsMA} is employed to automate this search process~\cite{ashouri_survey_2018}. Autotuning consists of applying various compiler optimizations on a program, evaluating their performance, and then choosing the best-performing option. Key challenges for autotuning include determining which optimizations to consider, which parameter to explore, and in what sequence to apply the optimizations~\cite{Basu2013TowardsMA}. Traditionally, autotuners relied on iterative compilation, where the code had to be repeatedly compiled and executed to assess the benefit of each optimization, making the procedure slow and computationally expensive. Modern autotuners instead employ predictive models, i.e., machine learning models that act as fast surrogates, providing performance predictions without iterative compilation~\cite{ashouri_survey_2018, ml-in-compiler-optimization}. In this work, we integrate supervised machine learning within the autotuning framework to estimate effective combinations of transpilation passes for a given quantum circuit.

The workflow for supervised machine learning consists of three stages: engineering the \explVars{}, training the model, and deploying the model~\cite{kampezidou-fundamental-supervised-ml}.
The success of a machine learning  project is largely determined by the choice of \explVars{}, sometimes also called features~\cite{domingos-a-few-useful-things-about-ml}. These are quantifiable and meaningful attributes of the underlying problem that act as inputs to machine learning models.
In the context of this paper, \explVars{} could, for instance, encode properties of quantum circuit intermediate representations, and are typically expressed as integer, real-valued, or boolean quantities.    

The learning process can be broken down into three core elements~\cite{domingos-a-few-useful-things-about-ml}: (1) a decision component that generates a labeled dataset by mapping input data to either a continuous output value or a discrete class label, along with an associated error or loss function; (2) an evaluation procedure that measures how effectively the model performs on previously unseen data; and (3) an iterative training phase that updates the model’s tunable parameters to minimize error and thereby optimize performance.
During deployment, the iteratively trained model is put into use: end-users or systems submit their problem instance, the corresponding \explVars{} are computed for that instance and fed into the model, which then produces the prediction.

%% file: sections/03_overview.tex
\section{Overview and Preliminaries of our Autotuning Approach}

In this section, we present an overview and the necessary preliminaries of our contribution.
To improve the effectiveness of optimizations performed in a transpiler pipeline, we propose a methodology in which machine-learning predictor is trained on a new dataset, that we generated by applying optimization passes to a diverse collection of quantum circuits. The concrete goal for the optimizations proposed in this paper is to reduce the number of two-qubit gates, which are the highest cause for noise in \gls{nisq} quantum devices.
Thus, the reduction of two-qubit gates is often used in our reasoning, and later also a major focus in our evaluation.
A major advantage of our approach is that it enables the automatic selection of effective optimization passes for individual circuits, without demanding extensive domain expertise and remaining device-agnostic. 

\subsection{Methodology}

A visualized overview of our complete methodology, which we call an autotuning pipeline, is shown in \autoref{fig:methodology}. Our contribution is divided into the following six steps:
\begin{figure}[b]
    \centering
    \includegraphics[width=\columnwidth]{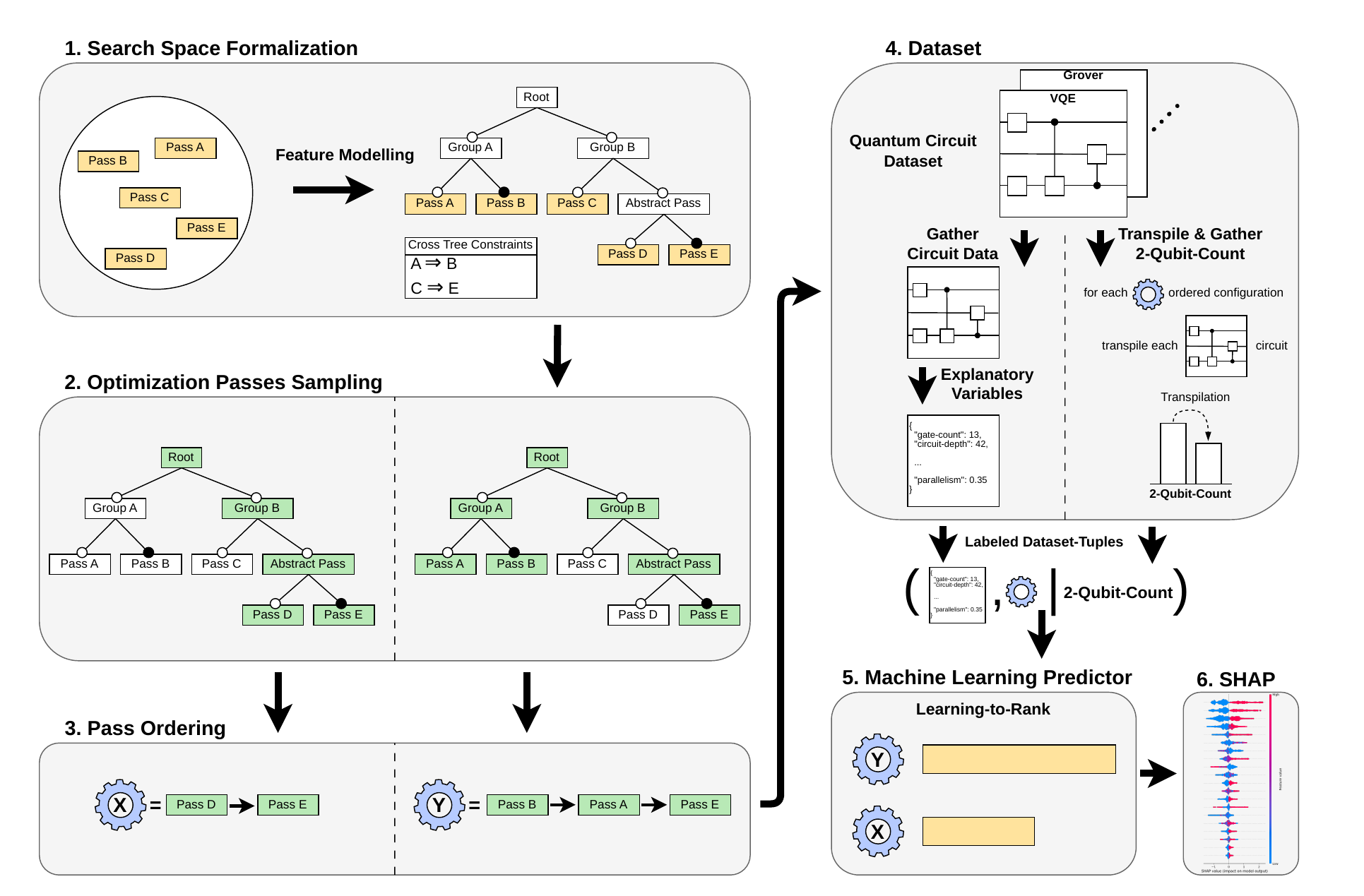}
    \caption{Overview of our autotuning pipeline. This figures includes all six steps that we performed to create this contribution, including the creation of a dataset via a formalization and sampling and pass ordering (1-4), the training of a machine learning predictor (5), and an evaluation of our results via a \gls{shap} analysis (6).}
    \label{fig:methodology}
    \Description{Shown is a Figure that gives an overview of the Methodology that is presented in this paper. Presented are 6 stages including a 1. Search Space Formalization that is illustrated as a feature model, 2. Optimization Pass Sampling which is also illustrated as a feature model with selected features, 3. Pass ordering which is illustrated as features / transpiler passes that are put in an order, 4. a dataset creation based on quantum circuits and feature model configurations, 5. a machine learning predictor that ranks configurations, and 6. a figure created by the SHAP analysis.}
\end{figure}

\begin{enumerate}
    \item \textbf{Search Space Formalization:} To propose suitable optimization passes, we must first search for and understand how different combinations of these passes operate for a particular class of quantum circuits. The Qiskit transpiler pipeline, for instance, offers 33 optimization passes for both init and optimization stages. All 33 passes can theoretically be flexibly combined, leading to a combinatorial explosion in the search space. However, some optimization passes are abstract and cannot be directly applied to a circuit, and many passes are interdependent (certain passes must be executed together because one generates results required by another). Consequently, naively assembling passes without considering such dependencies can result in invalid combinations of optimization passes. To address this, we first formalize the space of valid combinations of optimization passes using feature modeling, which captures variability through features and their relationships~\cite{kang1990feature} and allows us to encode selectable passes together with their dependencies expressed as cross-tree constraints. This approach formalizes the search space as the set of all combinations of optimization passes that satisfy the feature model (shown in yellow in \autoref{fig:methodology}).
    
    \item \textbf{Optimization Passes Sampling:} The feature model formalization captures the search space, but does not mitigate its combinatorial explosion. To mitigate the combinatorial explosion, we use the feature model as the foundation for sampling, a method that reduces the search space to a representative subset of combinations of optimization passes. Given our feature model and cross-tree constraints, sampling selects passes to form sampled combinations of optimization passes, denoted as configurations (shown in green in \autoref{fig:methodology}).

    \item \textbf{Pass Ordering:} Transpiler pipelines are highly sensitive to how passes are ordered. Consequently, executing the same set of passes in different sequences can yield different results. The order in which passes need to be executed is something that our feature model, nor our configurations, do not yet express. To tackle this, we introduce an \textit{Execution Order}, which specifies a concrete ordering that can be applied to configurations, illustrated as numbered blue cogs in ~\autoref{fig:methodology}.

    \item \textbf{Dataset:} The ordered configurations now represent concrete combinations of optimization passes, together with the sequence in which they are to be applied to a quantum circuit. We next apply these ordered configurations to a variety of quantum circuit classes from the quantum benchmarking suite \mqtBench{}~\cite{quetschlich2023mqtbench} to quantify how effectively each ordered configuration reduces the two-qubit gate count. The resulting reduction serves as a label indicating the performance of each ordered configuration on each circuit. We additionally compute \explVars{} that extract the characteristics of the circuits and together with configurations can serve as suitable input for a machine learning predictor, which learns for quantum circuit which configurations reduce the two-qubit gate count the most. We combine these labeling and explanatory-variable extraction steps into an automated data generation pipeline, which produces a labeled dataset tailored for training the machine learning predictor.

    \item \textbf{Machine Learning Predictor:} The labeled dataset captures how different quantum circuits respond to each ordered sample of optimization passes. Our aim is to generalize these observations so that, after training, the predictor can be used to predict suitable ordered configurations for previously unseen circuits without actually executing the ordered configurations. To this end, we train our machine learning model in a \learningToRank{} setting~\cite{learning-to-rank-paper-nair}: given a new circuit expressed with \explVars{}, the model produces an two-qubit-gate count reduction ranking over all candidate ordered configurations found during \textbf{Optimization Passes Sampling} step, effectively predicting which ordered combination of optimization passes is most likely to yield the largest reduction in two-qubit gate count. 

    \item \textbf{SHAP:} While our machine learning predictor serves as a fast surrogate that, for a given circuit, recommends ordered samples of optimization passes expected to yield the largest reduction in two-qubit gates, its internal decision process remains opaque: it is not directly clear why a particular ranking is produced. In addition to accurate predictions, we therefore aim to explain why specific optimization passes tend to enhance (blue) or degrade (red) performance, measured in two-qubit-gate reduction, for different classes of circuits, as illustrated in ~\autoref{fig:methodology}. To this end, we employ \gls{shap}~\cite{shap}, an interpretability framework that quantifies the contribution of each explanatory variable and each chosen optimization pass, aggregated over circuit classes, to the rankings produced by our predictive model. Such insights can inform the design of more fine-grained strategies for selecting optimization passes in future work.
\end{enumerate}

\subsection{Application Scenario}

Our contribution is not only purely methodological but also implemented for the actual quantum transpilation pipeline from Qiskit. When describing the details of our autotuning approach, we try to keep explanations abstract so that they can be understood without knowing the details of Qiskit's implementation, but sometimes reasoning with actual implementation details is necessary to explain certain design decisions. Below we motivate the scope of the application scenario in our implementation.

As described in \autoref{background:qiskit-subsec}, the Qiskit transpiler pipeline consists of six stages, two of which, the init and optimization stages, are dedicated to circuit optimization, primarily by reducing the number of two-qubit gates, which we use as our primary performance metric. In this work, we restrict our attention to optimizations applied in the init stage. We impose this restriction because, in Qiskit’s default transpiler pipeline, by the time a circuit reaches the optimization stage, it has already been adapted to a specific device’s topology, connectivity constraints, and \gls{isa}. In contrast, init-stage optimizations operate on device-agnostic circuits and are therefore unaffected by device-specific biases. By focusing on the init stage, we can attribute any reductions in two-qubit gate count directly to the optimizations themselves and to our methodology for choosing them. This, in turn, provides insights that future work could extend to later stages of the pipeline and to scenarios in which the target hardware is selected.

\subsection{Optimization Target}
\label{subsec:optimization-target}

As described in \autoref{sec:conept}, our primary metric is the reduction in the number of two-qubit gates, which is only well-defined once the circuit is expressed in terms of explicit, concrete gates. However, the optimizations in the init stage are \gls{isa}- and hardware-agnostic, and they allow the circuits to be represented in any form suitable for transpilation. In such representations, two-qubit gates may be concealed within more abstract constructs, reflecting how circuits are typically specified before entering the transpilation pipeline. To evaluate our metric, the reduction of two-qubit gates, in a way that is both meaningful and realistic, we must therefore introduce a translation stage to a concrete target gate basis before running the optimizations, in order to record the baseline for comparison, and again after the optimizations have completed, to quantify the improvement. We choose the following target gate basis for this translation:
\begin{equation}
\label{eq:target-basis}
\mathcal{B} = \{R_z, SX, X, CX\}.
\end{equation}
The basis $\mathcal{B}$ is a universal quantum gate set, meaning that any quantum circuit can be decomposed into circuits using only these gates~\cite{abughanem2026early}. For this to be possible, the gate set must be capable of generating all quantum superpositions as well as quantum entanglement.
To do this, the basis $\mathcal{B}$ includes the RZ, SX, and X gates. These gates implement rotations around the Z and X axes by fixed angles. The RZ gate supports rotations by any chosen angle, whereas the SX gate performs only $90^{\circ}$ rotations and the X gate performs $180^{\circ}$ rotations, which corresponds to flipping the qubit state (a bit flip). Other superposition-generating gates, such as the Hadamard gate, can be constructed using combinations of these basic gates. Quantum entanglement is provided by the CX gate, which acts on two qubits, one serving as the control and the other as the target. When the control qubit is in the state $\ket{1}$, the gate flips the state of the target qubit. In this way, it can create entanglement between the two qubits.

Our choice for this specific gate set is further motivated by the fact that it matches the native \gls{isa} of current state-of-the-art IBM superconducting quantum processors~\cite{abughanem2026early}. Moreover, this basis is in line with the evaluation typically used in quantum transpiler research~\cite{quetschlich2023compiler, quetschlich2023predicting, quetschlich2024mqt_predictor}, enabling us to evaluate our research against previous contributions with similar goals.

%% file: sections/04_concept.tex
\section{Applying Autotuning to Quantum Transpiler Pipelines}
\label{sec:conept}

With the scope defined, and thus a clear focus on a specific segment of the Qiskit transpilation pipeline, we now describe the six phases of our autotuning approach (see~\autoref{fig:methodology}) in detail. 

\input{sections/04_01_formalization}

\input{sections/04_02_sampling}
\input{sections/04_03_ordering}
\input{sections/04_04_dataset}

\input{sections/04_05_machinelearning}

\input{sections/04_06_SHAP}

%% file: sections/04_01_formalization.tex
\subsection{Search Space Formalization}
\label{subsec:search-space-formalization}

The first step of our autotuning pipeline is a formalization that captures the variability of quantum transpiler pipelines, which, in our case, need to capture possible combinations of Qiskit's optimization passes. The formalization later allows us to analyze how various combinations of optimization passes influence the circuits and gives us a systematic method to represent \textit{valid} combinations of passes.
In this context, valid means that passes that depend on each other are always selected together and that we avoid the selection of passes that are mutually exclusive.
Formalizing this validity is necessary because we want to avoid that our autotuning approach predicts incorrect combinations. 
Additionally, the formal representation gives us the possibility to automatically generate valid combinations of passes, which will be essential in later parts of our contribution, especially in the sampling and the data set creation.

To create the formalization, we decided to use a feature modeling approach~\cite{kang1990feature}, which allows us to capture the variability of the Qiskit transpiler pipeline within a feature diagram and additional cross-tree constraints. 
Feature models have the advantage that their formalization is visually appealing and easy to understand.
Additionally, they can be transformed into boolean satisfiability problems, allowing us to perform SAT-based analysis and sampling algorithms on them~\cite{eichhorn2023quantum}.
Formally, a feature model can be defined as:

\begin{definition}[Feature Model]
A feature model $\text{FM} = (\text{F}, \text{D})$ is a tuple that consists of a set of features $\text{F} = \{f_1, \dots, f_n\}$ and a set of dependencies $\text{D} = \{d_1, \dots, d_m\}$.
In the context of this paper, features correspond to transpiler passes, and dependencies correspond to propositional formulas that express the relations between them.
\end{definition}

An important aspect of feature models is the creation of so-called configurations, which are instances of a feature model where features are either selected or not selected. 
In our context, we want to map a configuration to a set of transpiler passes that we select to create a transpiler pipeline. 
Formally, we can define a configuration as follows:

\begin{definition}[Configuration]
\label{def:valid-configuration}
A configuration $c$ is a function that maps a set of features F to 0 (not selected) or 1 (selected):
\begin{center}
$c: \text{F} \rightarrow \{0,1\}, \quad \text{so that} \quad c(f_i) = 
\begin{cases}
    1, f_i \text{ is selected} \\ 0, f_i \text{ not selected}
\end{cases}$
\end{center}
We call a configuration \textit{valid} iff it satisfies all dependencies D. We denote the set of valid configurations as:
\begin{center}
$\mathcal{C}(\text{F}) = \{c \mid c \text{ satisfies } \text{D}\}$.
\end{center}
\end{definition}

\begin{figure}[ht]
    \centering
    \includegraphics[width=\textwidth]{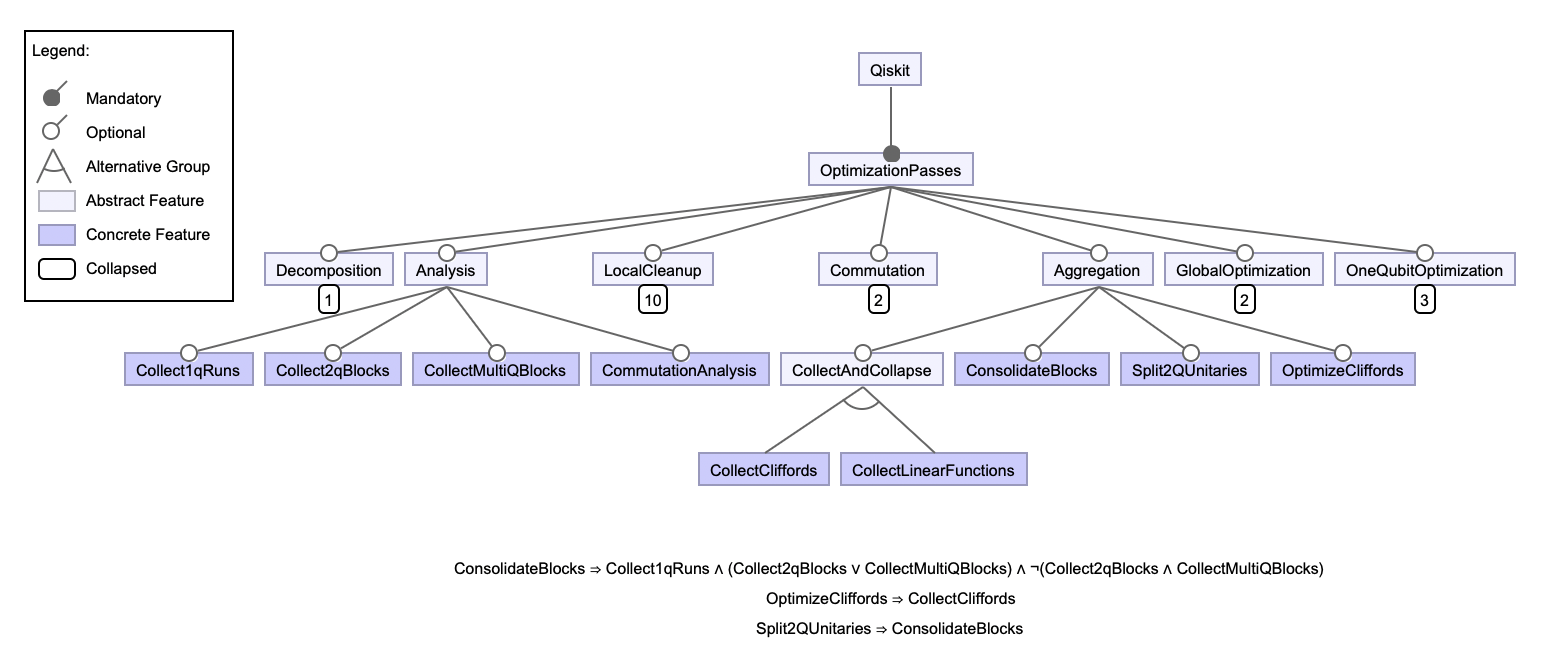}
    \caption{A feature model that represents the configurability of the transpiler passes appropriate for Qiskit's init stage.
    To improve the readability of the figure we had to collapse some nodes of features, a complete figure as well as a complete expression of the feature model in the Universal Variability Language can be found in our supplementary material~\cite{supplementary_material}.}
    \label{fig:feature-model}
    \Description{The figure shows the feature model that was used to formalize the search space of the qiskit transpiler pipeline. The feature model consists of 8 relevant groups and 27 features. Not all features are shown (collapsed) due to space limitations.}
\end{figure}

Based on this formal framework, we can now proceed to formalize the configurability of the Qiskit transpiler pipeline. The outcome of this process is presented in \autoref{fig:feature-model} and shows a comprehensive feature model that formalizes the Qiskit transpiler passes appropriate for the init stage. 

The core aspect that our formalization, and therefore the feature model, has to describe, is the complete set of transpiler passes available for our context.
Note that in our subject system, we only consider optimization passes and a single basis-change pass appropriate for the init stage. Therefore, our feature model excludes any passes that apply transformations for \gls{qpu}-specific layouts. To create this formalization, we used the information available in the official Qiskit documentation~\cite{qiskit_documentation_23}, and intensively studied the code of the open source implementation of the Qiskit transpiler pipeline from Qiskit version 2.3.0~\cite{qiskit_version_230}.
To denote all relevant passes, our feature model contains one \textit{concrete feature} for each transpilation pass. Thus, every such concrete feature corresponds to exactly one transpiler pass. 
To ensure our formalization captures only optimizations relevant to our subject system and remains compatible with modern \gls{nisq} devices, we restrict it to concrete passes that optimize circuits so that translation to the basis $\mathcal{B}$ defined in \autoref{subsec:optimization-target} accurately reflects the reduction in two-qubit gates.

In addition, the feature model includes several \textit{abstract features} that capture further logical aspects.  
These additional aspects describe, for example, compiler pass groups, which combine passes that pursue related objectives.  
Concretely, we define eight such groups: \textit{OptimizationPasses}, \textit{Analysis}, \textit{LocalCleanup}, \textit{Commutation}, \textit{Aggregation}, \textit{GlobalOptimization}, \textit{OneQubitOptimization} and \textit{Decomposition}.  
Each of these categories aggregates passes that share a common high-level optimization goal, even though they may realize this goal through different techniques.

So far, our formalization denotes and groups the available passes, but it not yet covers which valid pass combinations can be created.
To add this validity aspect, we now need to ensure that our feature model covers the dependencies that are associated between these passes. This can be achieved by using the structure of the feature diagram, and by defining additional cross-tree constraints to encode dependencies that cannot be covered by the feature diagram.
The feature diagram gives us the following possibilities to express dependencies:
\textit{Mandatory} indicates that a child feature is included in every configuration that contains its parent feature; \textit{Optional} indicates that the child feature may or may not be present in a configuration when the parent is selected; \textit{Alternative Group} specifies that exactly one of the child features can be included in a configuration if the parent feature is present; and \textit{Or} requires that, if the parent is selected, at least one of its child features must be included in the configuration. As shown in \autoref{fig:feature-model}, we define a single mandatory relationship, namely between the root feature \textit{Qiskit} and the logical group \textit{OptimizationPasses}. The abstract root feature \textit{Qiskit} represents the overall system under study. The mandatory relation to \textit{OptimizationPasses} narrows the scope of our model to the optimization passes offered by Qiskit, meaning that every configuration is derived exclusively from the optimization-pass search space. Furthermore, the model contains only one \textit{Alternative Group}, linking the general feature \textit{CollectAndCollapse} to its two concrete realizations, \textit{CollectCliffords} and \textit{CollectLinearFunctions}. These two features represent different implementations of the same abstract operation, but encode mutually exclusive strategies.
The remainder of the model is composed of \textit{Optional} relationships. Although some passes could in principle be modeled as mandatory children of others to capture internal dependencies, we avoid doing so in order to preserve the grouping of the features we introduced, rather than implementation-level dependencies between passes. Using mandatory-child relationships for this purpose would compromise the intended meaning of these logical groupings, so we instead encode most dependencies as cross-tree constraints.

To express dependencies without using the hierarchy of the feature diagram, feature models also allow the definition of additional cross-tree constraints. These cross-tree constraints are expressed as additional propositional formulas. In \autoref{fig:feature-model} they are listed below the feature diagram.
In our case, we use the cross-tree constraints to express (a) transformation passes that depend on specific analysis passes, and (b) transformation passes that must be preceded by other transformation passes that first rewrite gate-level circuits into more abstract forms. This is crucial for modeling optimization in Qiskit, where we identified three such dependencies. All three are \textit{Requires} cross-tree constraints, meaning that the selection of one feature mandates the inclusion of another. 
Following we give one example where we describe the reasoning for the constraint in detail:

\begin{example}
\textit{ConsolidateBlocks} is a transformation pass that replaces sequences of consecutive gates with a Unitary object, which is an abstract representation of these gate blocks as a unitary matrix rather than as gate instructions. This abstraction facilitates re-synthesis aimed at minimizing the number of 2-qubit gates. \textit{ConsolidateBlocks} consumes the precomputed block-list and run-list data and clears both once it completes. The block-list can originate from either of the analysis passes \textit{Collect2qBlocks} or \textit{CollectMultiQBlocks}. Since they both write to the same shared buffer under the block-list, the two passes cannot be applied together in any valid configuration. The run-list is generated by the \textit{Collect1qRuns} pass. Formally, this can be expressed as follows:
\begin{equation}
\label{eq:consolidate-blocks-constraint}
\text{ConsolidateBlocks} \Rightarrow \text{Collect1qRuns} \land (\text{Collect2qBlocks} \oplus \text{CollectMultiQBlocks})
\end{equation}
Adding this constraint to the feature model ensures that, in every valid configuration, whenever consolidation is enabled, each of the required buffers is written to exactly once. 
\end{example}

%% file: sections/04_02_sampling.tex
\subsection{Configuration Sampling}
\label{subsec:config-sampling}

Although the formalized feature model representation of the search space captures dependencies between optimization passes, it alone does not yet resolve the combinatorial explosion present in the search space. Exhaustively evaluating every valid configuration is computationally infeasible, and manually deciding, for each circuit, which passes are likely to be advantageous is time-consuming and usually limited to domain experts. Therefore, autotuning focuses on exploring only a representative subset of the overall search space~\cite{ashouri_survey_2018}. Exploration of this subset can be done with sampling, which refers to a process of selecting a subset of candidate configurations from the search space whose performance, in our case the reduction of 2-qubit gates, will be evaluated~\cite{auto-tuning-in-high-performance-computing}.

In traditional compiler optimization~\cite{ashouri_survey_2018}, the search space is usually explored via iterative compilation, which employs search heuristics such as random sampling, genetic algorithms, or simulated annealing~\cite{iterative-compilation-random-sampling, iterative-compilation-genetic-algorithms-sampling, iterative-compilation-simulated-annealing-sampling}. Although these iterative compilation strategies can mitigate the combinatorial explosion of the search space, they do not account for dependencies and constraints. As our methodology with feature models enables capturing the variability of transpilation pipelines, and thus the dependencies and constraints, we aim to use an appropriate sampling strategy that explicitly takes these defined variabilities into account.

To tackle this challenge, we employ a \textit{t}-wise interaction sampling strategy, where the objective is to cover all \textit{t}-wise interactions between selectable features using as few configurations as possible~\cite{strategies-for-testing-spl-intro-t-wise}. When applied to our application scenario where the search space is represented as a feature model, \textit{t}-wise interaction sampling guarantees that every valid \textit{t}-way combination of features, i.e., optimization passes, occurs in at least one configuration. This ensures that specific combinations of optimization passes, whose interactions may be advantageous or harmful, are systematically represented in the sampled configurations. To apply sampling to the feature model, we use the YASA algorithm that was introduced by Krieter et al.~\cite{krieter_yasa_2020} and is considered a state-of-the-art approach~\cite{UnWise}. Using this approach, our goal is to generate a Configuration Set $\mathcal{S}$:
\begin{definition}[Configuration Set]
\label{def:sample}
A configuration set $\mathcal{S} \subseteq C(F)$ is a set of sampled configurations $\{c_1,\dots,c_n\}$, where each configuration is valid according to \autoref{def:valid-configuration} and each transpiler pass corresponds to a \textit{concrete feature} described in \autoref{subsec:search-space-formalization}.
\end{definition}
To construct $\mathcal{S}$, YASA systematically enumerates all \textit{t}-wise feature interactions, uses SAT solving to filter out invalid interactions, and incrementally builds a compact set of valid configurations that cover all remaining interactions. Because it relies on SAT solving, YASA ensures that every selected configuration adheres to the feature-tree hierarchies and cross-tree constraints defined in our formalized feature model, while omitting \textit{abstract features}, since these cannot be applied to a transpilation pipeline. Empirical studies suggest that \textit{2}- or \textit{3}-wise interaction sampling is typically sufficient to expose most relevant interactions~\cite{krieter_yasa_2020, variabilty-qualitative-analysis, burgstallerOptimizationSpaceLearning2024}. In this work, given the size of our dataset and our available computational budget, we use \textit{3}-wise interaction sampling to generate a larger set of configurations, enabling us to uncover more interactions than a \textit{2}-wise strategy would.

%% file: sections/04_03_ordering.tex
\subsection{Pass Ordering}
\label{sec:pass_ordering}
With the set of sampled configurations $\mathcal{S} \subseteq C(F)$ established, we can now tackle the ordering problem by introducing an \emph{\execOrder{}}. 
Defining such an \execOrder{} is essential because quantum transpiler pipelines depend on the sequence in which their passes are applied. Consequently, applying the same collection of passes in different sequences can lead to different outcomes. 
As per \autoref{def:valid-configuration}, $C(F)$ currently denotes only a set, implying that $\mathcal{S}$ specifies which passes are applied to circuits, but not the order in which they are executed.
Consequently, we must extend the current definition of configurations with the notion of ordering. To achieve this, we introduce the following \execOrder{} that has to be applied to every $c \in \mathcal{S}$:

\begin{definition}[Execution Order]
Given a configuration $c \in \mathcal{S}$ with a set of transpilation passes $\{p_1,\dots,p_n\}$ that are selected in $c$ (meaning that $c(p) = 1$), then the \textit{\execOrder{}} $\phi$ of $c$ is defined as a sequence $\phi_c = [p_{\sigma(1)},\dots,p_{\sigma(n)}]$, where $\sigma$ is a permutation of the selected transpiler passes that is determined by applying the following rules:
\begin{enumerate}
    \item \textbf{Strategy Rule:} transpilation passes that follow shared objectives should be grouped and the groups executed in a order that maximizes optimization potential.
    \item \textbf{Dependency Chain Rule:} transpilation passes that depend on each other must execute sequentially, immediately one after another.
    \item \textbf{Mandatory Translation Rule:} final occurrence of a transpilation pass that creates or works on abstract representations, not suitable for subsequent transpilation passes, must be followed by a corresponding translation to a suitable representation.
\end{enumerate}
\end{definition}
To define the rules that create the \execOrder{}, in the following, we refer to the specific transpilation passes involved. However, a detailed understanding of each individual transpiler pass is not required in order to grasp why these rules are needed.

The \strategyRule{} addresses problems that arise when deciding the order in which to apply passes contained in logical groups from our feature model. If these logically grouped passes, with each group reflecting a distinct underlying optimization goal, are executed in an arbitrary sequence, the resulting order of groups, and thus of their passes, can lead to different outcomes. In particular, it may cause lost optimization opportunities, because some objectives are best pursued early in the circuit optimization process, while others are more effectively addressed at later stages. The \strategyRule{} solves this issue by prescribing a fixed ordering for all logical groups. To construct this ordering for our application scenario, we drew inspiration from how Qiskit arranges its optimization passes during the init stage~\cite{qiskit_version_230}. Considering both the function of optimization passes selected by Qiskit and their ordering, we derive an analogous ordering for our own methodology, where the concrete ordering enforced by the \strategyRule{} is defined as follows:

\begin{definition}[Strategy Rule]
\label{def:strategy-rule}
Passes are executed based on their category, in the order:
\begin{align*}
&\text{1. Decomposition}
\rightarrow \text{2. Analysis}
\rightarrow \text{3. LocalCleanup}
\rightarrow \text{4. Commutation}
\rightarrow \\&\text{5. Aggregation}
\rightarrow \text{6. GlobalOptimization}
\rightarrow \text{7. OneQubitOptimization}.
\end{align*}
Inside each of these categories, passes are arranged according to a simple lexicographic ordering.
\end{definition}

The \dependencyChainRule{} applies to situations in which one transpilation pass generates an intermediate artifact (such as a buffer or a modified circuit via abstract objects) that a later pass relies on. In these cases, the intermediate artifact must not be overwritten, ensuring that each pass works with a valid intermediate result. In our application scenario, the Qiskit transpiler pipeline, this problem occurs for sequences of passes that are mutually dependent, as specified by cross-tree constraints. The \dependencyChainRule{} solves this issue by enforcing a strict, sequential \execOrder{} on any passes that appear together in a given constraint.

\begin{definition}[Dependency Chain Rule]
\label{def:dependency-chain-rule}
~
\begin{enumerate}
    \item ConsolidateBlocks, when enabled, shall be executed immediately after the analysis passes Collect1qRuns and either Collect2qBlocks or CollectMultiQBlocks.
    \item Split2QUnitaries, when enabled, shall be executed immediately after ConsolidateBlocks.
    \item OptimizeCliffords, when enabled, shall be executed immediately after CollectCliffords.
\end{enumerate}
\end{definition}

Lastly, the \mandatoryTranslationRule{} covers issues that arise from transpilation passes that transform parts of a circuit into abstract representations that subsequent passes might be unable to interpret. 

The \mandatoryTranslationRule{} avoids possible issues by ensuring that the final occurrence of any pass that generates an abstract representation, or final occurrence of any pass that consumes these representations, must be directly followed by a matching translation pass that converts the circuit back to a representation expressed in quantum gates.
In our application scenario, this rule can be implemented by enforcing the ordering based on the following definition:

\begin{definition}[Mandatory Translation Rule]
\label{def:mandatory-translation-rule}
~
\begin{enumerate}
    \item If ConsolidateBlocks is selected and Split2QUnitaries is not selected in a configuration, then ConsolidateBlocks must be immediately followed by UnitarySynthesis.
    \item If Split2QUnitaries is selected in a configuration, then Split2QUnitaries must be immediately followed by UnitarySynthesis.
    \item If CollectCliffords is selected and OptimizeCliffords is not selected in a configuration, then CollectCliffords must be immediately followed by HighLevelSynthesis.
    \item If OptimizeCliffords is selected in a configuration, then OptimizeCliffords must be immediately followed by HighLevelSynthesis.
    \item If CollectLinearFunctions is enabled in a configuration, it must always be immediately followed by HighLevelSynthesis.
\end{enumerate}

\end{definition}

With the ordering rules in place, we can now apply the \execOrder{} to our configuration set $\mathcal{S}$. This yields an ordered configuration set $\mathcal{S}_{\text{ordered}}$, which we can then use to construct the data set required to train our machine learning predictor.

\begin{definition}[Ordered Configuration Set]
\label{def:ordered_samples}
The ordered configuration set $\mathcal{S}_{\text{ordered}}$ is derived by obtaining the \execOrder{} $\phi$ for every configuration $c$ of the configuration set $\mathcal{S}$, i.e.
$\mathcal{S}_{\text{ordered}} = \{\phi_c \mid c \in S\}$.

\end{definition}

%% file: sections/04_04_dataset.tex
\subsection{Dataset}
\label{subsec:dataset}

After applying the \execOrder{} to our configuration set and deriving $\mathcal{S}_{ordered}$, we can now start creating the dataset necessary to train the machine learning predictor of our autotuning approach. 

In the following, we explain how we use our ordered configurations to construct a large-scale dataset that allows us to empirically analyze how various optimizations passes influence different quantum circuits, evaluated according to our optimization objective, the reduction in two-qubit gates. Furthermore, we determine the \explVars{} that characterize the underlying quantum circuits in a representation suitable for our machine learning predictor. 

A very important aspect when creating the dataset is to make it sufficiently large and diverse, because different classes of quantum computing problems may respond differently to the same optimizations.
To achieve this goal, we obtain quantum circuits from the benchmark library \mqtBench{}~\cite{quetschlich2023mqtbench}.
The \mqtBench{} library contains thousands of quantum circuits spanning 26 different categories of quantum computing problems and corresponding algorithms including \gls{qaoa}~\cite{farhi2014quantumapproximateoptimizationalgorithm}, \gls{vqe}~\cite{kandala2017hardware}, and \gls{qnn}~\cite{kwak2021quantumneuralnetworksconcepts}, each offered at multiple abstraction levels tailored to various quantum tools and use cases. 
Here, the chosen abstraction level limits where a machine-learning predictor can be applied: If a model is trained on circuits already mapped to a particular device, it can only generate predictions at that specific level of abstraction.
To match our intended application scenario, which is improving Qiskit’s init-stage optimization, we therefore need to restrict our dataset to abstraction level that correspond to the \emph{Target-Independent Level}. 
Following this choice, we obtain a total of 1943 distinct circuits from \mqtBench{} that can be investigated without relying on any hardware-specific details.

The next important aspect for the creation of our dataset is the definition of a metric that corresponds to our optimization objective, the reduction of two qubit gates, which we use to measure how well a specific combination of transpilation passes (e.g., one configuration of $\mathcal{S}_{ordered}$) performed.
This metric is subsequently used to guide the machine-learning predictor so that, for any given circuit, it prioritizes those ordered configurations that are most likely to produce the largest decrease in two-qubit gates. More precisely, the metric compares the number of two-qubit gates in a circuit after the ordered configurations from $\mathcal{S}_{ordered}$ have been applied to the baseline two-qubit gate count obtained before any optimization. 
Both the pre- and post-optimization two-qubit gate counts are evaluated in the basis $\mathcal{B}$, as justified in \autoref{subsec:optimization-target}. 
In the following, we refer to this metric as the \emph{Optimization Ratio} and define it as follows:

\begin{definition}[Optimization Ratio]\label{def:optimization-ratio}
Let $b(q_i)$ be the baseline two-qubit gate count of a circuit $q_i$ obtained by translating $q_i$ into the basis $\mathcal{B}$ and counting its two-qubit gates and let $o(q_i,\phi_c)$ be the optimized two-qubit gate count, obtained by applying the ordered configuration $\phi_c \in \mathcal{S}_{ordered}$ to $q_i$, translating the result into $\mathcal{B}$, and counting its two-qubit gates. We define the \emph{Optimization Ratio} as: $$\text{optRatio}(q_i, \phi_{c}) = \frac{b(q_i)}{o(q_i,\phi_c)},$$ where $\text{optRatio}(q_i, \phi_{c}) > 1$ indicates a reduction in two-qubit gates, with higher values corresponding to higher reductions.
\end{definition}

To train our machine learning predictor on the relationship between a circuit and its resulting \emph{Optimization Ratio}, we now need a suitable representation of each circuit, denoted as \emph{\explVars{}} and motivated in \autoref{subsec:bacckground_compiler_autotuning_and_supervised_ml}. 
Thus, we associate every quantum circuit with a vector of \emph{\explVars{}} for a given circuit $q_i$ denoted with $\text{Exp}_i$, defined as follows:
\begin{definition}[Explanatory Variables]
\label{def:\explVars{}}
For each circuit, the \emph{\explVars{}} vector is defined as a 62-dimensional vector constructed from the following components:
\begin{enumerate}
    \item 13 \explVars{} obtained using the Qiskit circuits properties, describing scalar circuit properties such as depth, width, and size~\cite{qiskit_documentation_23},
    \item 42 \explVars{} counting the occurrences of standard gates in the \emph{Target-Independent Level} (e.g., $CX$) as specified in the circuit's OpenQASM 2 description~\cite{cross2017openquantumassemblylanguage},
    \item 5 \explVars{} obtained from the SuperMarQ benchmarking framework~\cite{tomesh_supermarq_2022}, namely Program Communication, Critical Depth, Entanglement Ratio, Parallelism, and Liveness,
    \item 2 \explVars{} we constructed ourselves: the number of gates acting on exactly 2 qubits and the number of gates acting on exactly 3 qubits.
\end{enumerate}
A complete listing of all \explVars{} is provided in our supplementary material~\cite{supplementary_material}.
\end{definition}

In summary, to construct the dataset for our application scenario, we apply every ordered configuration $\phi_c \in \mathcal{S}_{ordered}$ to each of the 1943 quantum circuits obtained from \mqtBench{}, record the resulting Optimization Ratio, and compute the corresponding \explVars{}. This procedure yields all the information required to train our machine-learning predictor. As this dataset may also be useful for future related research and reproducibility, we make it publicly available in our supplementary material~\cite{supplementary_material}.

%% file: sections/04_05_machinelearning.tex
\subsection{Surrogate Machine Learning}
\label{subsec:surrogate-ml}
\begin{figure}[h]
    \centering
    \includegraphics[width=\columnwidth]{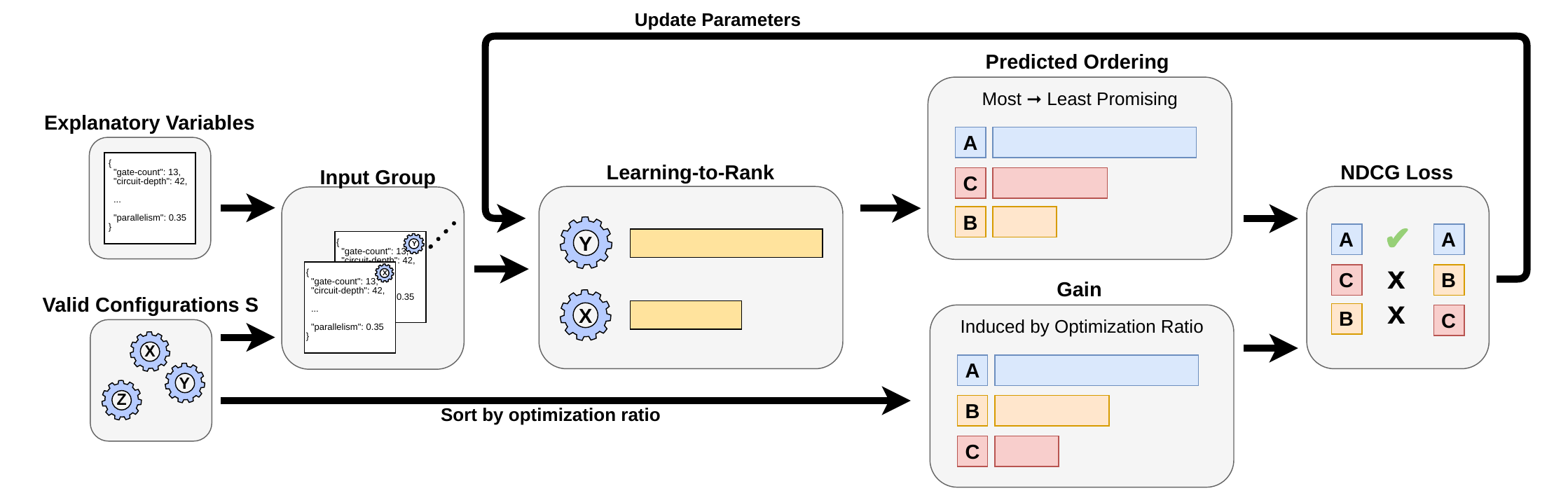}
    \caption{Overview of our machine learning predictor formulated as a \learningToRank{}~task. This figure presents a simplified illustration of the model input, the resulting ranking of configurations, and the way the model adjusts its parameters using a loss function.}
    \label{fig:ml-pipeline}
    \Description{The figure shows an overview of the machine learning pipeline implemented in the paper. This figure presents a simplified illustration of the model input, the resulting ranking of configurations, and the way the model adjusts its parameters using a loss function.}
\end{figure}

With the labeled dataset established, we can specify our machine learning predictor with a supervised machine learning algorithm. Nair et al.~\cite{learning-to-rank-paper-nair} showed that compiler optimization is best framed as selecting the best configuration from a candidate set via ranking, rather than predicting a single configuration. Motivated by this, we model our machine learning predictor as a supervised \learningToRank{} problem with the XGBoost machine learning algorithm~\cite{xgboost-paper}. Given a quantum circuit, this predictor induces a ranking for the configurations $c \in \mathcal{S}$ from most to least promising. The highest-ranked configuration can then be selected to achieve the largest reduction in the 2-qubit gate count of the quantum circuit. In the following, we describe how each step of the machine learning works following the visualizations in~\autoref{fig:ml-pipeline}.
First, our predictor requires an input that enables it to rank configurations from most to least promising, as depicted by the yellow bars in~\autoref{fig:ml-pipeline}. The \emph{\explVars{}} defined in~\autoref{def:\explVars{}} describe the circuit and are identical for every configuration, thus they alone do not allow to rank configuration from a pool of candidates. To support the ranking process, we additionally inform the machine learning predictor which configurations are currently being assessed for a given circuit, forming \emph{Input Groups}:
\begin{definition}[Input Group]\label{definition:ml-input-group} For a given circuit $q_i$, we define the grouping of all configurations $c \in \mathcal{S}$ as
$$\mathcal{I}_i = \{(Exp_i,c) \mid c \in \mathcal{S}\},$$
i.e., the set of configuration pairs that share the \explVars{} vector $Exp_i$ of circuit $q_i$. Note that each pair contains a binary configuration $c \in \mathcal{S}$ rather than its ordered counterpart $\phi_c \in \mathcal{S}_{ordered}$, since the predictor treats a configuration as a binary unordered selection of passes. Each pair is  processed independently of each other in parallel by the predictor.
\end{definition}
With the \emph{Input Group} specified, we now turn to the central aspect of any machine learning predictor: its output. As outlined above, each element of the \emph{Input Group} is processed independently and in parallel. Consequently, the predictor cannot directly produce a ranking in the sense of a joint enumeration over integer values for all available configurations, since each prediction is made in isolation and the model has no explicit knowledge of the full set of candidate configurations. Instead, for each input pair $(Exp_i, c) \in \mathcal{I}_i$, our machine learning predictor outputs a \emph{Real-Valued Score} denoted by $v_{i,c} \in \mathbb{R}$. These scores can then be used to sort all configurations and thereby derive a \emph{Predicted Ordering}:
\begin{definition}[Predicted Ordering] We define the \emph{Predicted Ordering} for a circuit $q_i$ as a sequence of all configurations $c \in \mathcal{S}$: 
$$\varphi_i = [c_{\sigma(1)},\dots,c_{\sigma(k)}],$$ where $\sigma$ is a permutation of the configurations determined by their \emph{Real-Valued Score}, such that: $v_{i,c_{\sigma(1)}} \geq v_{i,c_{\sigma(2)}} \geq \dots \geq v_{i,c_{\sigma(k)}}$, where $k = \lvert \mathcal{S} \rvert$ and ties are broken deterministically first by number of active passes and then by the configuration index. Thus, the place of a configuration in $\varphi_i$ encodes its predicted rank, with $c_{\sigma(1)}$ denoting the most promising configuration for circuit $q_i$.
\end{definition}
Subsequently, the model must be trained by updating its parameters, to refine the \emph{Predicted Ordering} described above, via a loss function. As our goal is identifying the best-performing optimization configurations, rather than enforcing a precise ordering among clearly suboptimal ones, among loss functions available to \learningToRank{}, we choose the loss based on \gls{ndcg}~\cite{ndcg-lambdaLoss}. This choice is motivated by the fact that \gls{ndcg} assigns stronger penalties to misrankings in \emph{Predicted Ordering} that affect the most beneficial optimization configurations for a given circuit, i.e., those that achieve the largest reduction in two-qubit gates. \gls{ndcg} does this by comparing the \emph{Predicted Ordering} of a \emph{Input Group} against a set of \emph{Gains} induced by \emph{Optimization Ratio}, quantifying how valuable each configuration is and, thus, how severe it is to misplace it in the ordering.

With the training concluded, we can employ our machine learning predictor to predict suitable optimizations for unseen circuits. For an unseen circuit, we compute the \explVars{} and configurations pairs as defined in \autoref{definition:ml-input-group}, run all the pairs through the machine learning predictor in parallel, obtaining one \emph{Real-Valued Score} per circuit-configuration pair, sort the score and select the top-ranked configuration which can be then be applied to a circuit in the Qiskit transpilation pipeline to obtain an optimized circuit.

%% file: sections/04_06_SHAP.tex
\subsection{SHAP Analysis}
\label{subsec:shap}
Beyond accurate prediction with our \learningToRank{} model, our goal is to understand why particular optimization passes help or hinder the reduction of two-qubit gates across different classes of circuits. One of the main challenges motivating our autotuning approach is the complexity of the underlying problem. Different classes of quantum algorithms can respond very differently to the same transpiler passes, and choosing an appropriate set of passes for a given circuit often requires deep expertise in both the problem structure and the circuit optimization techniques available. Our predictor learns these relationships empirically from data, but its internal decision process remains opaque due to its black-box architecture~\cite{shap}.

\gls{shap} provides the explainability in two ways. First, via Shapley values, it explains how each individual explanatory variable or applied optimization pass contributes to the predicted score, by quantifying its marginal effect relative to the predictors average output~\cite{molnar2025}. This indicates which aspects drive higher or lower ranking. Second, via the Shapley interaction index, \gls{shap} captures how pairs of \explVars{} or passes interact, revealing synergies: (a) where two elements together correlate more than expected from their separate effects or (b) antagonistic behavior, where one element suppresses the benefit of another~\cite{molnar2025}. These interaction analyses reveal which concrete circuit properties affect the effectiveness of an optimization pass, offering data-driven guidance for assembling optimization configurations. 

In this work, we perform this analysis by providing our trained predictor to the \gls{shap}~\cite{shap} framework, which evaluates the inputs drawn from the training set, and then decomposes the predictor’s output into contributions from individual Shapley values and also into Shapley interaction effects. The results of this analysis are provided and discussed in \autoref{subsec:explainability}.

%% file: sections/05_evaluation.tex
\section{Evaluation}
\label{sec:evaluation}
In this section, we empirically evaluate our proposed autotuning methodology using the Qiskit transpiler as an application scenario. 

We will focus on the extend of the improvements that we were able to achieve, compare the performance of our approach on different circuit classes, and explain the behavior of our predictor by employing interpretability frameworks.
To contextualize our results, we compare our machine-learning-based predictor with the heuristic optimization levels currently used in Qiskit's init stage, and with the \mqtPred{}~\cite{quetschlich2024mqt_predictor}, which is another well-established open-source framework that can perform automated transpiler pass selections.
Specifically, we address the following research questions:
\begin{enumerate}
    \item \textbf{Predictability}: Given \explVars{} that characterize a quantum circuit, how accurately can our predictor select, from all sampled configurations, the configuration that minimizes the circuit's two-qubit gate count the most?
    \item \textbf{Reduction Improvements}: How does the two-qubit gate reduction achieved by the predicted configuration compare, both overall and per algorithm class, to (i) Qiskit's init-stage optimization levels and (ii) a well-established reinforcement learning configuration selector \mqtPred{}?
    \item \textbf{Explainability}: Which selected optimization passes drive the two-qubit reductions, for which algorithm classes, and how can this knowledge be transferred into guidance for improving quantum transpilation pipelines?
\end{enumerate}
To answer these research questions, we first describe our experimental setup, then present the results and analyze them in a detailed discussion.

\input{sections/05_01_experiment_design}
\input{sections/05_02_results}

\subsection{Threats To Validity}

In the following, we discuss internal and external threats that might impact the validity of the methodology and the evaluation that we presented.

\subsubsection*{Internal Threats}

Our method requires an explicit specification of the \execOrder{} of passes. Consequently, our reported results are dependent on the particular \execOrder{} we selected, and changes in this ordering could impact the performance of our predictor and thus the conclusions drawn in our evaluation.
We mitigated potential threats to validity arising from this circumstance by deriving our ordering from a detailed analysis of the Qiskit transpiler pipeline implementation. 
As a result, the outcomes presented in our evaluation are more directly comparable, and the high performance of our predictor provides convincing evidence that the chosen ordering is appropriate.

Our machine learning predictor relies on a corpus of quantum circuits intended to represent a broad range of quantum computing applications. The validity of our claims about generalization and the identification of effective optimization passes via \gls{shap} depends on two key assumptions: (1) that this corpus adequately covers all major problem classes, and (2) that we have appropriately addressed potential overfitting and bias in the model training process. We mitigated this threat by using the well-established \mqtBench{} benchmarking suite, which spans 26 categories of quantum computing problems, and by following machine learning best practices for data handling, including stratified, group-aware train–test splits tailored to the \learningToRank{} formulation.

\subsubsection*{External Threats}

Our experiments are conducted on a hardware-agnostic, noiseless simulator to isolate the effects of the optimization passes themselves. However, this environment may not fully capture the behavior of real quantum devices, which are constrained by limited qubit connectivity and affected by noise. Consequently, it is uncertain to what extent our results carry over to later stages in the Qiskit compilation pipeline that adapt circuits to specific quantum hardware. Nevertheless, our methodology, together with the explainability results from our \gls{shap} analysis and our comparison against Qiskit init-stage optimizations in the same hardware-agnostic setting, provides valuable insights for this research area.

Our methodology is transferable to other quantum transpilers, but the concrete artifacts produced in this work (e.g. feature model, \execOrder{}, and trained predictor) are tailored to the specific Qiskit version used in our experiments. They depend on the available passes and their implementation in that version, so changes in Qiskit (e.g., new or modified passes or defaults) may limit their direct applicability and require retraining or adaptation. We reduced the risk of limited transferability by using Qiskit, currently the most influential quantum programming framework~\cite{upadhyay2025analyzing}, thereby ensuring that a broad segment of the quantum-computing community can benefit from our artifacts. Furthermore, we reduced the risk by publishing our full, reproducible experimental setup as open source with extensive supplementary documentation to support adaptation to future versions and other scenarios.

The results we report are evaluated using a specific basis $\mathcal{B}$, corresponding to the universal quantum gate set of IBM’s current superconducting quantum processors. This is only one of several gate sets compatible with the \glspl{isa} of existing quantum processors, which means that the artifacts produced by our methodology are closely tied to this particular basis. This creates a threat to the generalization of our findings. We mitigated this risk by choosing a basis that is widely used in quantum transpiler research~\cite{quetschlich2023compiler, quetschlich2023predicting, quetschlich2024mqt_predictor} and by designing our open-source experimental pipeline so that the underlying basis can be easily adapted to other gate sets and platforms.

%% file: sections/05_01_experiment_design.tex
\subsection{Experiment Design}
\label{subsec:experiment-design}

We evaluate our autotuning approach using a Docker-based environment. 
Reproducibility is ensured through straightforward commands that execute the respective pipeline stages with automatic persistence and the ability to resume interrupted runs. In this subsection, we describe the experimental setup and give an overview of the artifacts that were created during the steps described in our methodology.
We also make the entire experimental pipeline publicly available in our supplementary material~\cite{supplementary_material}.

\subsubsection*{Search Space Formalization}
The feature model introduced in \autoref{subsec:search-space-formalization} was defined in XML using FeatureIDE~\cite{FeatureIDE}. It contains 37 features in total, 27 of which correspond to concrete transpilation passes that can be applied to quantum circuits.

\subsubsection*{Configuration Sampling}
Using this feature model, configuration sampling with YASA, as described in \autoref{subsec:config-sampling}, was performed with the FeatJAR framework~\cite{featJAR}. Applying YASA to the feature model produced $62$ configurations in total, where each configuration consists of up to 19 transpiler passes, with an average of approximately $10.9$ applied passes per configuration.

\subsubsection*{Dataset}
The training circuits were sourced from \mqtBench{}~\cite{quetschlich2023mqtbench} version 2.2.2, resulting in a collection of 1943 quantum circuits spanning 26 categories of quantum computing tasks, with circuit sizes ranging from 2 to 127 qubits at the \emph{Target-Independent Level}. The circuits and configuration pairs described in \autoref{subsec:dataset} were transpiled using the Qiskit Transpilation Pipeline~\cite{qiskit_version_230} from Qiskit version 2.3.0 on a system equipped with 128 GB of RAM and an AMD Ryzen Threadripper PRO 5955WX CPU featuring 16 physical and 32 logical cores. Due to the varying sizes of the \mqtBench{} circuits and the configurations, we imposed a timeout of 120 seconds for each circuit–configuration pair. In total, we evaluated $120466$ circuit–configuration pairs, of which $109603$ finished in time.

\subsubsection*{Surrogate Machine Learning}
Using the resulting labeled dataset, we trained our prediction model described in \autoref{subsec:surrogate-ml} with the XGBoost~\cite{xgboost-paper} machine learning algorithm. To reliably evaluate how well the trained model generalizes to entirely unseen circuits we adopt a group-aware train-test scheme. Concretely, we divide the available circuits such that no specific circuit–configuration pair is present in both the training and test sets within a single split. Furthermore, we apply group-aware stratification over quantum computing problems, as defined in \mqtBench{}~\cite{quetschlich2023mqtbench} to guarantee that, for each group, some circuits are included in both the training and evaluation phases. In each split, roughly 80\% of the circuits from all groups are allocated to training, with the remaining 20\% reserved for testing. To tune the XGBoost machine learning model, we use the Optuna~\cite{akiba2019optunanextgenerationhyperparameteroptimization} framework, which automatically searches for hyperparameters that maximize model performance. The hyperparameter search used a GroupKFold cross-validation scheme provided by sklearn~\cite{scikit-learn}. The resulting values and trained model are available in our supplementary material~\cite{supplementary_material}.

\subsubsection*{SHAP}
For the \gls{shap} analysis described in \autoref{subsec:shap}, we applied the TreeExplainer algorithms from the \gls{shap} library~\cite{shap} to the trained XGBoost model. Because the computation of Shapley interaction values scales quadratically with the number of elements defined in our \emph{Input Group} (\autoref{definition:ml-input-group}), we sampled uniformly 500 datapoints for the interaction analysis. For the standard Shapley values, which scale linearly in the number of elements in the \emph{Input Group}, we conducted the analysis on the complete set of available data-points.

%% file: sections/05_02_results.tex
\subsection{Predictability}
To answer the first research question, we examine whether given a circuit and sampled configurations, our predictor can correctly rank these configurations so as to select the one that removes the largest number of two-qubit gates for that circuit. To this end, we report the following metrics:
\begin{enumerate}
    \item \textbf{Top 1 and Top 3 Accuracy:} Rate with which the best found sampled configuration is within the the top-1 or top-3 configurations the predictor proposes.
    \item \textbf{Mean Regret and Median Regret:} Mean and Median difference in the \emph{Optimization Ratio} achieved by our predictors proposed top-1 configuration, compared to sampled configuration that yields the highest \emph{Optimization Ratio} for the given circuit.
\end{enumerate}
\begin{table}[H]
  \centering
  \caption{Accuracy of the predictor.}
  \label{tab:ranker_accuracy}
  \begin{tabular}{cccc}
    \toprule
    Mean Regret & Median Regret & Top-1 Acc. & Top-3 Acc. \\
    \midrule
    $3.4 \times 10^{-5}$ & 0.0 & 98.7\% & 99.5\% \\
    \bottomrule
  \end{tabular}
\end{table}
Table~\ref{tab:ranker_accuracy} summarizes the performance of our machine-learning-based predictor on the 20\% test split, resulting in 391 test-circuits, stratified by the class of quantum algorithms. 
On average across all test-circuits, our predictor selects the Top 1 configuration in $98.7\%$ of the cases out of the 62 configurations defined in \autoref{subsec:experiment-design}, and in $99.5\%$ of the cases its recommended configuration lies within the Top 3. The strength of our predictor is further highlighted by the Mean Regret and Median Regret values: whenever the our predictor does not select the best found configuration, the average missed optimization potential relative to the best configuration is negligible, and the median missed potential is zero.

\subsubsection*{\textbf{Answering RQ1}}
The results demonstrate that the predictor can very accurately predict the top-1 sampled configurations that achieve the largest reduction in two-qubit gate counts for a given circuit. In the few cases where our predictor does not identify the best sampled configuration, the resulting loss in optimization potential is negligible, indicating that the model has effectively learned from the dataset. Thus, end-users can rely on the ranking proposed by our predictor.

\subsection{Reduction Improvements}\label{subsec:reduction-improvements}
To address the second research question, we evaluate whether using our predictor is indeed beneficial compared to Qiskit’s built-in init-stage optimization levels and the well-established \mqtPred{}, with respect to the achieved  two-qubit gate reductions. 

When comparing ourselves against Qiskit, we compare against the following three optimization levels:
(O0) at optimization level 0, Qiskit applies no optimization passes; (O1) At optimization level 1, Qiskit applies two transpiler passes in the following order: \textit{InverseCancellation} and then\textit{ContractIdleWiresInControlFlow};
(O2) At optimization level 2, Qiskit applies the following transpiler passes in the order: 
\begin{align*}
    &\text{Unroll3qOrMore}
    \rightarrow \text{RemoveDiagonalGatesBeforeMeasure}
    \rightarrow \text{RemoveIdentityEquivalent}
    \rightarrow \\&\text{InverseCancellation}
    \rightarrow \text{ContractIdleWiresInControlFlow}
    \rightarrow \text{CommutativeCancellation}
    \rightarrow \\&\text{ConsolidateBlocks}
    \rightarrow \text{Split2QUnitaries.}
\end{align*} 
(O3) Optimization level 3 is currently identical to optimization level 2. We decided to still include the results in the evaluation to avoid intransparency.

When comparing ourselves with the \mqtPred{} reinforcement-learning approach~\cite{quetschlich2024mqt_predictor} (RL), which we adapted to operate on the same search space of optimization passes as defined in \autoref{subsec:search-space-formalization} and to use the same two-qubit gate reduction metric as described in \autoref{subsec:dataset}.

\begin{figure}[h]
    \centering
    \includegraphics[width=\columnwidth]{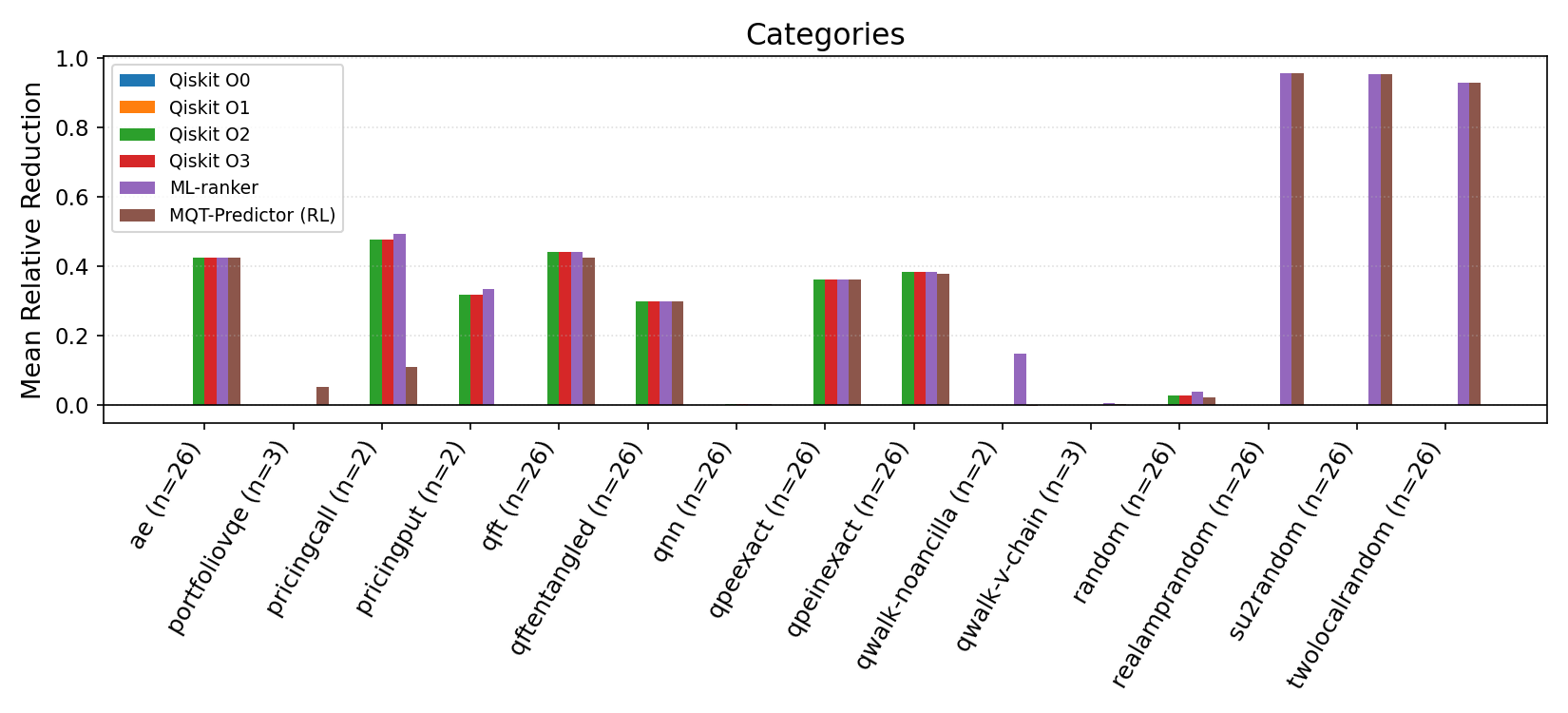}
    \caption{Mean Relative Reduction of two-qubit gates across quantum circuit classes. Of the 26 total classes, only the 15 classes for which at least one method achieved a reduction are shown. The number of evaluated circuits for each class is denoted with n.}
    \label{fig:per-category-reduction}
    \Description{A figure that shows the Mean Relative Reduction of two-qubit gates across quantum circuit classes. Of the 26 total classes, only the 15 classes for which at least one method achieved a reduction are shown.}
\end{figure}

For a fair, quantitative comparison, we report results for the following metrics:
\begin{itemize}
    \item \textbf{Wins, Draws, and Losses:} The number of test-set circuits for which the configuration predicted by our predictor achieves a better, equal, or worse reduction in two-qubit gates compared to the respective Qiskit optimization level or reinforcement-learning baseline.
    \item \textbf{Mean Relative Reduction and Median Relative Reduction:} The mean and median fractions of removed two-qubit gates relative to the corresponding baseline.
    \item \textbf{Mean Absolute Reduction and Median Absolute Reduction:} The mean and median absolute differences in the number of removed two-qubit gates relative to the corresponding baseline.
\end{itemize}
\begin{table}[h]
  \centering
  \caption{Head-to-head comparison of our predictor against Qiskit init-stage optimization levels O0--O3 and the \mqtPred{} (RL). Numbers are rounded to 1 decimal place.}
  \label{tab:headtohead_vs_qiskit_ranker}
  \resizebox{\columnwidth}{!}{%
    \begin{tabular}{rrrrrrrr}
        \toprule
        Approach & Wins & Draws & Losses & Mean Rel.\ Red. & Median Rel.\ Red. & Mean Abs.\ Red. & Median Abs.\ Red. \\
        \midrule
        O0 & 248 & 143 & 0 & 32.4\% & 1.5\% & 2695.6 & 42.0 \\
        O1 & 248 & 143 & 0 & 32.4\% & 1.5\% & 2695.6 & 42.0 \\
        O2 & 110 & 281 & 0 & 19.1\% & 0.0\% & 1640.9 & 0.0 \\
        O3 & 110 & 281 & 0 & 19.1\% & 0.0\% & 1640.9 & 0.0 \\
        RL & 55 & 329 & 7 & 0.7\% & 0.0\% & 27.6 & 0.0 \\
        \bottomrule
    \end{tabular}%
  }
\end{table}
As shown in Table~\ref{tab:headtohead_vs_qiskit_ranker}, across 391 test-circuits our predictor selects configurations that yield stronger two-qubit gate reductions than Qiskit’s light-weight optimization levels (O0, O1) for $63.4\%$ of the circuits and outperform the heavier optimization levels (O2, O3) for $28.1\%$ of the circuits, while never producing a configuration that performs worse on any test-circuit. Compared to \mqtPred{}, our predictor still performs favorably: it loses on only $1.8\%$ of the circuits, but overall obtains very similar two-qubit gate reductions on a large fraction of the test set. Considering the \textit{Mean Relative Reduction}, our approach removes on average $19.1$--$32.4\%$ more two-qubit gates than Qiskit’s optimization levels. However, the \textit{Median Relative Reduction} lies between $0$ and $1.5\%$, indicating that these large improvements are not uniformly distributed across all circuits, but are instead concentrated on a subset of them. To pinpoint where our predictor provides the most benefit, we therefore analyze, in \autoref{fig:per-category-reduction}, the fine-grained \textit{Mean Relative Reduction} over circuits from each of the 26 quantum algorithm categories contained in \mqtBench{}.

\autoref{fig:per-category-reduction} shows that the largest two-qubit gate reductions over Qiskit’s optimization levels are concentrated in three specific categories of quantum algorithms. Inspecting the \textit{Mean Relative Reduction} achieved by Qiskit’s O2 and O3 levels (green and red), we observe that for most categories their performance is comparable to that of our predictor and the \mqtPred{} reinforcement-learning approach. Notably, both our approach and the reinforcement learning approach consistently outperform Qiskit’s O0 and O1 levels (blue and orange). In contrast, for the \textit{realamprandom}, \textit{su2random}, and \textit{twolocalrandom} categories, our predictor and \mqtPred{} achieve $95.8\%, 95.4\%$ and $92.8\%$ mean reductions in two-qubit gates, whereas all Qiskit optimization levels achieve $0.0\%$. Furthermore, in the \textit{qwalk-noancilla} category, our machine-learning predictor identifies configurations that reduce two-qubit gate counts beyond what is achieved by both all Qiskit’s optimization levels and the reinforcement-learning-based \mqtPred{}.

\subsubsection*{\textbf{Answering RQ2}}
In summary, these findings demonstrate that our predictor is not only accurate but also clearly advantageous in terms of two-qubit gate reduction, making it a practical surrogate for Qiskit’s built-in init-stage optimization levels. Our predictor never selects configurations that lead to worse two-qubit gate counts than any of Qiskit’s optimization levels on any test-circuits, a trait the reinforcement learning agent does not provide. Crucially, our machine learning predictor delivers what Qiskit’s predefined init-stage optimization levels do not: targeted reductions of up to $95.8\%$ of two-qubit gates for specific classes of quantum algorithms by proposing circuit-specific optimization configurations. Thus, it can be reliably used in practice, even though for a substantial fraction of circuits the achieved improvements are relatively modest. When compared to the reinforcement learning \mqtPred{}, our approach is able to identify on some circuit classes stronger optimization configurations, although its overall impact on two-qubit gate reduction is much smaller than the gains observed relative to Qiskit’s optimization levels. 

In the following, we use the \gls{shap} explainability framework to make the optimization knowledge learned by our predictor persistent and transferable. Because our predictor evaluates every configuration for a given circuit, the \gls{shap} framework enables a global interpretation of its predictions across all circuit classes simultaneously, an approach that is not available for a reinforcement learning agent, which makes decisions individually and in a sequential fashion.

\subsection{Explainability}\label{subsec:explainability}
To answer the third research question, we now analyze which optimization passes, in combination with which circuit characteristics, drive the observed reductions in two-qubit gate counts. While the evaluation of the previous research questions has shown that our predictor is both highly accurate (RQ1 - Predictability) and capable of outperforming Qiskit and \mqtPred{} in two-qubit gate reductions (RQ2 - Reduction Improvements), the focus here is on making the learned optimization knowledge explicit and interpretable. We report explainability results obtained with the \gls{shap} framework using three types of plots:
\begin{itemize}
    \item \textbf{Mean SHAP Plot:} Shows the distribution of mean absolute Shapley values for selected optimization pass, thereby ranking them by their average influence on the predictor's predictions.
    \item \textbf{SHAP Beeswarm Plot:} Provides a comprehensive summary of the distribution of Shapley values for every
    selected optimization pass, highlighting how feature values affect the direction and magnitude of the prediction.
    \item \textbf{Cross-Group Interaction Plot:} Summarizes Shapley interaction values to show how circuits \explVars{} that characterize particular classes of quantum algorithms interact to drive the selection of optimization passes.
\end{itemize}
To connect directly to results for RQ2 (Reduction Improvements), we first present the \emph{Mean SHAP Plot} and \emph{SHAP Beeswarm Plot} for the \textit{realamprandom} class of quantum circuits, where our method achieves substantial reductions and Qiskit finds no optimizations.
\begin{figure}[htbp]
  \centering
  \begin{minipage}{0.5\textwidth}
    \centering
    \includegraphics[width=\linewidth]{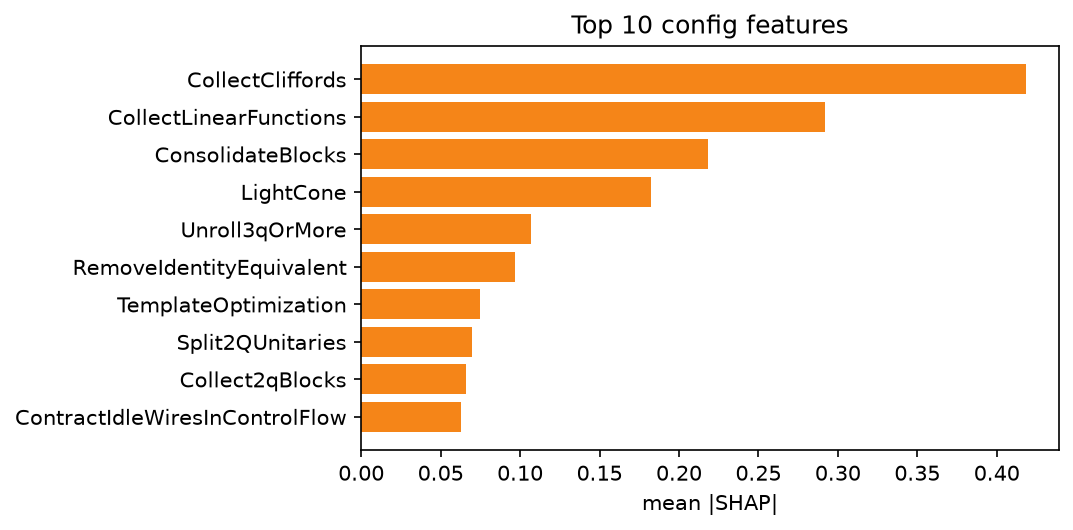}
  \end{minipage}\hfill
  \begin{minipage}{0.45\textwidth}
    \centering
    \includegraphics[width=\linewidth]{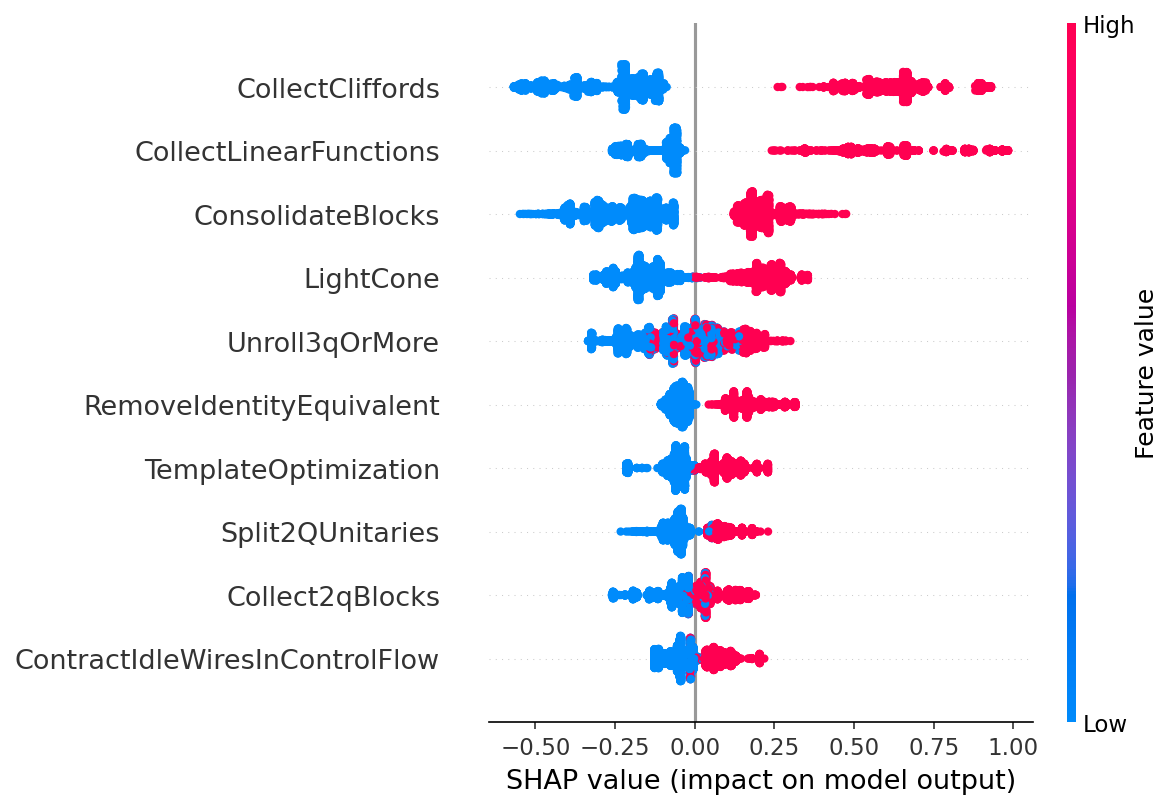}
  \end{minipage}
  \caption{Mean $\mid$\gls{shap}$\mid$ and \gls{shap} Beeswarm plots of Top 10 optimization passes influence for Real Amplitudes (Random) class of quantum computing problems.}
  \label{fig:shap-mean-realamprandom}
  \Description{Shown is a Figure that presents the results of the SHAP analysis in mean absolute SHAP and Beeswarm plots.}
\end{figure}
The left plot in \autoref{fig:shap-mean-realamprandom} reports the mean absolute Shapley values of the 10 most influential selected optimization passes for the \textit{realamprandom} circuit class. The most important passes are \textit{CollectCliffords} and \textit{CollectLinearFunctions}, with mean absolute Shapley values of approximately $0.42$ and $0.29$, respectively, clearly dominating the contribution to the predictor output. 
The right plot in \autoref{fig:shap-mean-realamprandom} further decomposes these contributions into their positive and negative impact on the predictor output. For \textit{CollectCliffords}, configurations in which this pass is absent (blue points) consistently exhibit negative Shapley values (approximately $-0.5$ to $-0.1$), indicating that the predictor is discouraged from selecting such configurations. In contrast, configurations that include \textit{CollectCliffords} (red points) show strictly positive Shapley values (approximately $0.25$ to $0.9$), meaning that the presence of this pass strongly pushes the predictor towards choosing the corresponding configuration. A similar, though slightly weaker, pattern is visible for \textit{CollectLinearFunctions}. Looking at all selected optimization passes in the right plot of \autoref{fig:shap-mean-realamprandom} reveal a consistent pattern: nearly all blue points are associated with negative Shapley values, whereas the red points correspond to positive Shapley values. This indicates, that choosing these optimization passes sends a strong signal to the predictor to select the corresponding configurations, as they correspond to a higher rank in the \textit{realamprandom} circuit classes. However, the strength of this effect varies in magnitude across the passes, with \textit{CollectCliffords} and \textit{CollectLinearFunctions} driving the reduction of the two-qubit gates.
\begin{figure}[h]
  \centering
  \begin{minipage}{0.45\textwidth}
    \centering
    \includegraphics[width=\linewidth]{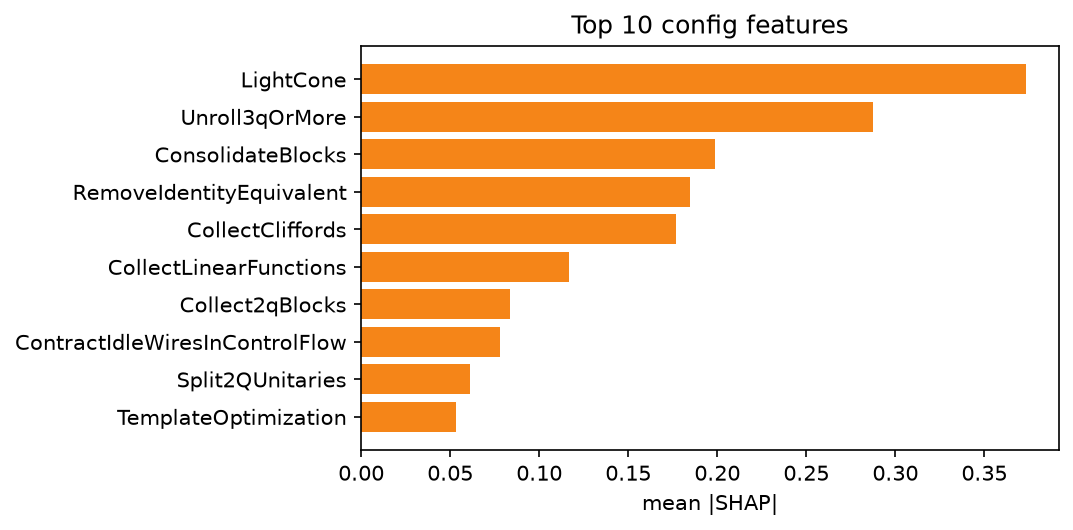}
  \end{minipage}\hfill
  \begin{minipage}{0.45\textwidth}
    \centering
    \includegraphics[width=\linewidth]{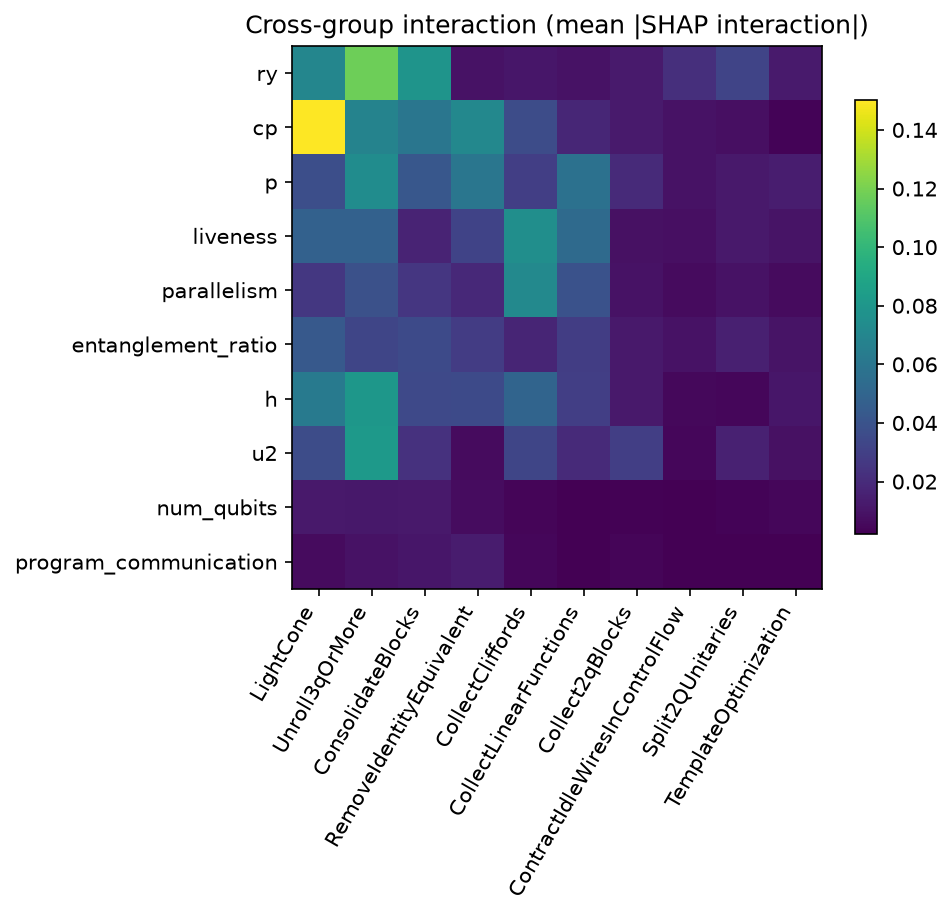}
  \end{minipage}
  \caption{The left plot shows mean absolute Shapley values displaying Top 10 optimization passes. Bigger mean $\mid$\gls{shap}$\mid$ corresponds to higher influence of the corresponding optimization pass on our predictor. The right plot shows a Cross-group Interaction Heatmap. Rows correspond to \explVars{} and columns correspond to optimization passes. Cells represent the magnitude of the mean absolute \gls{shap} interaction between a pass and explanatory-variable. Lighter color indicates stronger interaction. Sorted from left to right by the mean $\mid$\gls{shap}$\mid$ importance.}
  \label{fig:top-config-features-and-cross-interaction-heatmap}
  \Description{Shown is a figure that shows the results of the SHAP analysis as a cross-group interaction heatmap.}
\end{figure}

Another way to render the acquired optimization knowledge more explicit and interpretable is to examine how these selected optimization passes affect the predictor in the general case, by aggregating their influence across all circuit classes, as shown in the left plot of \autoref{fig:top-config-features-and-cross-interaction-heatmap}. Across this global view, \textit{LightCone} and \textit{Unroll3qOrMore} clearly dominate the predictor’s predictions, with mean absolute Shapley values of approximately $0.37$ and $0.29$, respectively. Other passes, such as \textit{TemplateOptimization}, also contribute to two-qubit gate reductions, but on a smaller scale, while still remaining within the Top-10 most influential passes.

However, these global importance scores alone do not reveal in which algorithmic contexts individual passes provide their strongest signal. Since, from the predictors’s perspective, circuit classes are distinguished only by their \textit{\explVars{}}, we analyze how pass importance depends on these variables using the cross-group interaction heatmap in the right plot of \autoref{fig:top-config-features-and-cross-interaction-heatmap}. Each cell reports the mean absolute \gls{shap} interaction between an optimization pass and an explanatory-variable, so reading reading columns from left to right shows which passes most affect the prediction overall, and reading a concrete cell in a column shows which circuit characteristics most strongly modulate the influence of a given pass. For example, \textit{LightCone} not only ranks as the most influential pass globally, but its effect is particularly amplified when circuits contain \textit{cp} or \textit{ry} gates, while still being fairly uniformly relevant across other \explVars{}. In contrast, the \textit{CollectCliffords} column shows weaker cumulative interaction values, with influence concentrated on a few specific variables such as liveness and parallelism. This aligns with the class-specific results shown in \autoref{fig:shap-mean-realamprandom}, and indicates that \textit{CollectCliffords} is a context-dependent pass that benefits only particular circuit families. Finally, the \textit{Split2QUnitaries} column exhibits no strong interactions with any explanatory-variable, suggesting that this pass rarely drives substantial two-qubit gate reductions and is not amplified by identifiable circuit characteristics. Together, these explanations make the learned optimization behaviour explicit, revealing which passes matter, in which quantum circuit classes, and how they jointly drive two-qubit gate count reductions.

\subsubsection*{\textbf{Answering RQ3}}
Taken together, applying \gls{shap} to our predictor yields persistent and transferable optimization knowledge. The results in \autoref{fig:shap-mean-realamprandom} clarify why our predictor achieves substantial two-qubit gate reductions on \textit{realamprandom} circuits where Qiskit’s default optimization levels do not: the most influential passes highlighted by our predictor, most notably \textit{CollectCliffords} and \textit{CollectLinearFunctions}, are absent from Qiskit’s init-stage optimization configurations, causing Qiskit to miss optimization opportunities that our predictor successfully uncovers. In our supplementary material~\cite{supplementary_material}, we provide mean \gls{shap} and beeswarm plots for all circuit classes, thereby pinpointing which specific passes, when applied to which circuit classes, drive the observed optimization gains. Subsequently, the results presented in the left plot of~\autoref{fig:top-config-features-and-cross-interaction-heatmap} reveal, aggregated across all circuit classes, the most influential optimization passes. This empirical knowledge can be used to refine heuristic approaches, such as Qiskit’s predefined optimization levels, by including highly effective passes such as \textit{CollectCliffords}, in order to recover optimization potential Qiskit currently misses on specific circuit classes. Finally, building on the identification of influential passes across all circuit classes, right plot of \autoref{fig:top-config-features-and-cross-interaction-heatmap} shows which easily computed \explVars{} amplify the relevance of specific optimization passes, laying the groundwork for designing even more sophisticated heuristic strategies.

%% file: sections/06_related_work.tex
\section{Related Work}

In this section we discuss how our work relates to previous works employing machine learning in the fields of highly configurable systems and quantum compilation.

\subsubsection*{\textbf{Configuration Prioritization and Configuration Performance Learning}}
Prior work in configuration performance learning uses various machine learning approaches on small samples of software configuration spaces (i.e., search spaces) to address their combinatorial explosion, by following a sampling, measuring, learning pattern~\cite{pereiraLearningSoftwareConfiguration2019}.
The aim of these models is to simplify the configuration of software systems by identifying configurations that perform optimally (i.e. prioritizing them over other configurations)~\cite{siegmundPerformanceinfluenceModelsHighly2015}.
For a detailed overview on configuration performance learning we refer to the surveys of \citeauthor{pereiraLearningSoftwareConfiguration2019}~\cite{pereiraLearningSoftwareConfiguration2019} and \citeauthor{gongDeepConfigurationPerformance2024}~\cite{gongDeepConfigurationPerformance2024}.

Several studies use feature-coverage heuristics for configuration performance prediction~\cite{pereiraLearningSoftwareConfiguration2019}.
Here, \textit{t}-wise sampling is commonly employed or compared against~\cite{yilmazCoveringArraysEfficient2006, siegmundPredictingPerformanceAutomated2012, sarkarCostEfficientSamplingPerformance2015, siegmundPerformanceinfluenceModelsHighly2015, kalteneckerDistanceBasedSamplingSoftware2019}. 

Furthermore, configuration performance prediction models are used to address the variability of compiler infrastructure~\cite{pereiraLearningSoftwareConfiguration2019}, making them a promising candidate also for quantum compilers.
In particular, LLVM is a commonly used subject system~\cite{siegmundPredictingPerformanceAutomated2012, guoVariabilityawarePerformancePrediction2013, zhangPerformancePredictionConfigurable2015, sarkarCostEfficientSamplingPerformance2015, siegmundPerformanceinfluenceModelsHighly2015, zuluagaEPALActiveLearning2016, ohFindingNearoptimalConfigurations2017a, kalteneckerDistanceBasedSamplingSoftware2019, kolesnikovTradeoffsModelingPerformance2019, nairFindingFasterConfigurations2020, Guething_2024}.
Compiler autotuning is explicitly addressed by Optimization Space Learning~\cite{burgstallerOptimizationSpaceLearning2024}, a supervised learning approach based on collaborative filtering shown to be a fast, noniterative method for the C compiler GCC.

\learningToRank{} approaches are often used in classical compiler optimization literature~\cite{ashouri_survey_2018}. A \learningToRank{} approach that motivated our choice of model was employed by \citeauthor{learning-to-rank-paper-nair}~\cite{learning-to-rank-paper-nair}, which showed that their model is cheaper to learn than residual-based configuration performance prediction models.
To the best of our knowledge, no previous work has used \learningToRank{} in a quantum compilation setting.

\subsubsection*{\textbf{Machine Learning for Quantum Compilation}}

Prior work on machine-learning applications for quantum compilation spans architecture-level decisions down to individual pass transformations.

At the highest level, machine learning has been used for circuit-architecture and execution-target decisions.
Prior work in this space either learns predictors or search policies for parameterized circuit structures, or predicts suitable hardware targets and target-dependent execution-quality estimates~\cite{he2023gsqas,he2023gnn_qas,meng2021mcts_qas,wang2022quantumnas,quetschlich2024mqt_predictor,hopfImprovingFiguresMerit2025,tudisco2025gnn_hardware_selection}.
These decisions fix the input circuits, execution targets, and cost functions that lower-level transpilation and optimization then operate on.

Once a circuit and target are fixed, machine learning has also been applied to individual transpilation subproblems and pass-level transformations.
These works learn constructive policies for unitary synthesis, layout, placement, mapping, routing, circuit rewriting, or diagram simplification using neural predictors, sequence models, reinforcement learning, graph neural networks, or search-assisted variants~\cite{swaddle2017generating,zhang2020topological,moro2021quantum,he2021variational,chen2024efficient,paler2023machine,fan2022placement,ren2024leveraging,pozzi2022routing,huang2022dear,tang2024alpharouter,amer2024initial_mapping,li2024quarl,riu2025zxrl,kremer2025practicalefficientquantumcircuit}.

At the compiler-flow level, existing methods primarily differ in whether they learn an online policy or an offline predictor over already evaluated compilation choices.
Online reinforcement-learning methods treat compilation as a sequential control problem.
\citeauthor{quetschlich2023compiler}~\cite{quetschlich2023compiler} model quantum compilation as a Markov decision process in which actions apply compilation passes or move between compilation states, enabling flows that combine passes from Qiskit and TKET.
\citeauthor{liu2025portable}~\cite{liu2025portable} propose a portable auto-tuning framework based on Double Dueling Deep Q-Networks for quantum compilation optimization.
TuniQ~\cite{hasanat2026tuniq} instead operates inside Qiskit's staged transpilation pipeline and learns a MaskablePPO policy that selects or skips passes across the initialization, layout, routing, translation, optimization, and cleanup stages, using action masks to enforce valid stage-dependent choices and reward shaping to propagate end-to-end compilation quality.
These methods learn policies whose training signal comes from interaction with a compilation environment.
This makes them well suited to adaptive pass ordering and cross-stage dependencies, but ties learning to repeated and computationally expensive environment rollouts and to the stability, and interpretability limitations of reinforcement-learning.
Our approach instead uses supervised learning on an offline dataset of scored pass combinations.
This fixes the training data and makes the relative performance of multiple candidate pass combinations available for each circuit.

Offline supervised approaches differ in how much of this autotuning data they retain.
Classification-based compiler prediction treats compilation-option selection as a single-label prediction problem and learns from the best observed option combination for each circuit~\cite{quetschlich2023predicting}.
This formulation is useful when the goal is to predict one high-level option tuple, but it compresses the evaluated candidate set to a single target label.
Thus, it does not use the relative ordering of non-optimal candidates, which can in principle be retained when the candidate configurations are evaluated offline.
Prior work on rank-based configuration selection argues that exact performance prediction is not necessary when the goal is to identify good configurations; it can be sufficient to train a model that ranks candidate configurations~\cite{learning-to-rank-paper-nair}.
Motivated by these observations, we formulate compiler-pass prediction as a supervised \learningToRank{} problem over valid ordered pass combinations.

%% file: sections/07_conclusion.tex
\section{Conclusion and Future Work}
\label{sec:conclusion}
In this work, we addressed an important challenge of current quantum compiler-pipeline optimization approaches, which typically rely on manually designed heuristics and developer intuition to decide which optimization passes to apply.
To overcome this, we proposed an autotuning-based methodology that uses supervised machine learning to automatically learn which combinations of optimization passes are most effective for a given quantum circuit. Specifically, we trained a predictor that, given a circuit and sampled configurations of optimization passes, ranks these configurations by their expected ability to reduce the main source of error on \gls{nisq} devices: the number of noisy two-qubit gates.

To implement our predictor, we made several contributions to the use of machine learning in quantum compilation pipelines. We analyzed and formalized Qiskit’s init-stage optimization search space, enabling smart sampling strategies that mitigate combinatorial-explosion of this space and produce a representative subset of valid optimization configurations. With valid optimization configurations, we carried out what was, to the best of our knowledge, the largest empirical study of the behavior and efficiency of the selected transpilation passes: we built a dataset by applying 62 different combinations of transpiler passes to 1,943 quantum circuits spanning several categories of quantum computing problems, recorded the resulting reductions in two-qubit gate counts, and thus created a self-contained artifact for future research in quantum compilation and autotuning. Using this dataset, we used state-of-the-art methods to train a machine learning predictor that correlates quantum circuit structure to the optimization potential of different configurations, enabling circuit-dependent configuration ranking.

The experimental results from our evaluation show that our predictor has a high accuracy and correctly selects the best sampled configuration (out of the 62 considered configurations) for $98.7\%$ of test-circuits. 
Additionally, the configurations it recommends \textit{never} yield a smaller two-qubit gate reduction than any of Qiskit’s heuristic optimization levels. On average over all test circuits, the predicted configurations achieve an additional $19.1\%$–$32.4\%$ relative reduction in two-qubit gates, and for circuit families where Qiskit’s preset optimization levels offer $0\%$ improvement, our approach reaches reductions of up to $95.8\%$. Compared to the well-established \mqtPred{} framework, our predictor generally finds similar or better configurations and achieves a comparable overall two-qubit gate reductions. 
However, the most informative results from our predictor stem from the explainability analysis that we performed. Thanks to the methodology that we chose in our implementation, we were able to perform a \gls{shap} analysis to explain why our predictor performs better than Qiskit’s optimization levels. In this analysis, we determined which passes have the greatest impact on configuration rankings, and thus offer the highest overall optimization potential, considering both (a) all quantum circuits collectively and (b) each individual class of quantum computing problems.
As a result, we were able to provide a novel and extensive data-driven study that gives empirical insights on the performance of different transpiler passes,  which could be used to refine heuristic strategies, including Qiskit’s predefined optimization levels.

As future work, we plan to integrate our predictor in our own production tools as well as try to create a publicly available extension of the Qiskit transpilation pipeline that allows to use our predictor directly within Qiskit.